# ReCoDe: A Data Reduction and Compression Description for High Throughput Time-Resolved Electron Microscopy

Abhik Datta[1,2], Kian Fong Ng[1,2], Balakrishnan Deepan[1,2], Melissa Ding[3], See Wee Chee[1,2,4], Yvonne Ban[1,2], Jian Shi[1,2], Duane Loh[1,2,4]

[1]Centre for BioImaging Sciences, National University of Singapore, Singapore 117557.
[2]Department of Biological Sciences, National University of Singapore, Singapore 117557.
[3]Department of Computer Science and Engineering, Ohio State University, Columbus, OH 43210, USA.
[4]Department of Physics, National University of Singapore, Singapore 117551.

## Abstract

Fast, direct electron detectors have significantly improved the spatio-temporal resolution of electron microscopy movies. Preserving both spatial and temporal resolution in extended observations, however, requires storing prohibitively large amounts of data. Here, we describe an efficient and flexible data reduction and compression scheme (ReCoDe) that retains both spatial and temporal resolution by preserving individual electron events. Running ReCoDe on a workstation we demonstrate on-the-fly reduction and compression of raw data streaming off a detector at 3 GB/s, for hours of uninterrupted data collection. The output was 100-fold smaller than the raw data and saved directly onto network-attached storage drives over a 10 GbE connection. We discuss calibration techniques that support electron detection and counting (e.g. estimate electron backscattering rates, false positive rates, and data compressibility), and novel data analysis methods enabled by ReCoDe (e.g. recalibration of data post acquisition, and accurate estimation of coincidence loss).

## Introduction

Fast, back-thinned direct electron detectors are rapidly transforming electron microscopy. These detectors ushered in a "resolution revolution" for electron cryo-microscopy (cryo-EM), and the prospect of seeing sub-millisecond dynamics for in-situ electron microscopy. These transformations are driven by three key factors: (1) improved detection efficiency, (2) shorter detector readout times to better resolve individual electron events, and (3) algorithms that translate these advances into improved spatial and temporal resolution. Whereas the first two factors have received considerable attention, it remains impractical for many existing algorithms to process the very large raw output produced by these movie-mode detectors. Fortunately, the useful information on these raw data are typically sparse, hence a suitable data reduction and compression scheme should allow us to fully reap the advantages offered by these detectors.

Nearly all the useful information in a single raw detector image is contained within "secondary electron puddles", each of which is digitized from the cloud of secondary charged particles formed in the wake of individual high energy electrons passing through the detector's sensor. While the size and shape of secondary electron puddles contain some information[1], localizing the entry point of the incident electron from its electron cloud already noticeably improves the spatial resolution of the image. To accurately localize these electron puddles they must be spatiotemporally well separated (by increasing the frame rate or reducing the incident electron flux), thereby reducing the so-called coincidence loss[2]. This separation creates a very high raw data load when acquiring images that add up to a desired accumulated electron dose. For example, the memory needed to store the incident electron entry points, in a low coincidence-loss image (~6%) acquired at 0.01 e/pixel/frame is approximately a hundredth that of the raw detector readout, with the remainder holding only thermal readout noise.

Currently, there are three popular options to manage the large raw data loads that a high-throughput electron detector generates. First, which is typical in cryo-EM, is to employ a higher internal frame rate on the detector for counting electrons at low coincidence loss, but add many of these frames together before they are stored to disk. The downside here is the loss of temporal resolution in the added, stored images. The second option is to reduce the total data acquisition time. Here, an experimenter may fill terabytes of local hard disk with raw data for ten minutes, then wait at least twice as long to offload this data to a larger networked drive before more data acquisition can proceed. The third option is to collect data at the maximum detector frame rate but only store the frames that contain significant information. However, this strategy only works at high dose rates where individual pre-selected frames still show sufficient contrast for the experimenter to judge whether to keep or discard them. At such high dose rates, the experimenter has to either sacrifice spatial resolution or be limited to atomic resolution only for radiation-hard samples.

None of these three options are ideal, especially since the vast majority of these high data loads are storing only the detector's thermal and readout noise. Furthermore, these options also limit us from using faster detectors[3] to study dynamics at even shorter timescales. Naturally, reducing and compressing the raw data would obviate the need to choose between these three compromising options. If we stored only electron arrival events, we can enjoy high temporal and spatial resolution, while continuously acquiring movies of dose-sensitive samples at very low dose rates for practically hours, uninterrupted.

To efficiently connect downstream processing algorithms to raw movie-mode detector data, we propose and implement a data reduction and compression scheme capable of file size reductions that are as high as 100x for realistic electron-counting scenarios. The output of this scheme is a file format known as ReCoDe (Reduced Compressed Description). For simplicity, we refer to the reduction compression scheme as the ReCoDe scheme (or simply ReCoDe when the context is clear). In this scheme, the original raw data is first reduced to keep only the information regarding identified electron puddles, which are then further compressed. The ReCoDe scheme permits four reduction levels and several different types of lossless compression algorithms, whose combinations are discussed in this work. We also show how data stored in the least lossy ReCoDe reduction level can be re-processed post-acquisition to remove detector artifacts, especially those owing to slow drifts in the thermal background. Moreover, storing data at this reduction level retains the puddle shape and intensity information. Through several use cases, we show the benefits of retaining this information. One of these is coincidence loss estimation, where we show that puddle shape information is essential for accurate estimation. We also develop methods for estimating the prevalence of back scattered electrons and estimating the false positive rates of electron events using this information. For the DE-16 detector we estimated the ratio of primary to backscattered electrons to be ~8.6.

The ReCoDe scheme is sufficiently parallelizable such that data streams from even the fastest current detectors can be reduced and compressed "on-the-fly" onto networked storage disks using only modest computing resources, provided the raw data can be accessed before it is written to disk. For instance, the raw data stream of a low dose experiment (0.8 e/pixel/s) collected on a DE-16 detector (~3.08 GB/s throughput) can be reduced, compressed by 10 Intel Xeon CPU cores, then written to network-attached storage devices via a modest 10 gigabit ethernet connection. Furthermore, the ReCoDe data format has been designed for fast sequential and random access, so that frames can be decompressed into the reduced representation on demand.

The ReCoDe format allows in-situ electron microscopy movies to retain the high dose fractionation and sub-millisecond time resolution while extending acquisition time from minutes to hours. While more experiments that require both high temporal and spatial resolution are emerging[4–6], acquiring such movies for long time scales remains expensive, and in many cases, infeasible. For perspective, a 4 TB hard drive only accommodates about 21 minutes of data collection at a DE-16 detector's maximum data rate of 3.08 GB/s. For the same hard drive, ReCoDe, owing to its very significant file size reduction, potentially allows hours of uninterrupted data collection for low dose-rate, electron-counted imaging without compromising on temporal resolution.

The reduced and compressed data has sufficiently small output size so that other programs can read the data available on disk for on-the-fly processing while ReCoDe is still writing new data to the disk. Typically frames are read-out from detectors to RAM (random access memory) then immediately written to high-speed SSDs (solid-state drives). However, even the fastest SSDs do not have sufficient bandwidth for the data on the disk to be simultaneously read by other programs while being written to. Absent fast simultaneous read-write, there is a critical lack of feedback in low dose rate, long time experiments. The experimenter is left blind in such situations, as individual frames do not have sufficient contrast and the frames available on disk cannot be read to produce a summed image with sufficient contrast. This manuscript describes an efficient implementation of ReCoDe that allows "on-the-fly" feedback without interrupting data acquisition for hours.

A prominent example that exploits fast detectors is motion-correction in TEM (transmission electron microscopy). Here, the imaging resolution is demonstrably improved when fast detectors fractionate the total electron dose on a sample onto a time series of micrographs that are individually corrected for relative dose-induced motion[7]. In fact, recent work suggests that using more efficient data representation to further increase dose fractionation, hence finer time resolution, can improve spatial resolution[8]. ReCoDe allows users to flexibly fractionate their doses from tens to thousands of frames per second for more precise temporal resolution and drift-correction where possible.

Going forward, the readout rates of CMOS detectors may increase to their internal megahertz clock rates[9], or even into the gigahertz regime[10]. This uptrend is troubling if one considers, by default, that a detector's raw data output rate increases linearly with its readout rate. However, because the ReCoDe format has very little storage overhead per frame, in principle, its processing and storage rate only scales with the total electron dose when the detector readout rate is fast enough to resolve individual electron puddles. Consequently, the ReCoDe output rate will not increase substantially with megahertz frame rates when the total electron dose is held constant.

As electron microscopy becomes increasingly reliant on larger datasets and more complex processing and analysis workflows, there is an ever-greater push for publications to include raw data necessary for others to validate and reproduce the analyses[11]. We show that for several publicly available EMPIAR[11] datasets with moderate to high dose rates, ranging from 0.5 to 5.0 electrons/pixel/frame, ReCoDe achieves 2-8x compression, outperforming existing compression approaches. Improved compression will make public archives like EMPIAR more accessible and increase their adoption, and encourage deposition of raw micrographs facilitating validation of the structures produced using them[12]. Without an effective data reduction and compression scheme, storing raw detector data will be costly: at ~US$20 per terabyte (TB) of archival storage on commodity HDDs and ~US$400 per TB on SSDs (based on the prices of lower end external hard disk drives and solid-state drives, as of April 2019[13,14], just 15 minutes of continuous data acquisition per day on the DE-16 (Direct Electron, LP) detector at its maximum frame rate (3 GB/s throughput) will cost between US$ 20,000 to US$ 400,000 per year respectively.

Overall, we hope that by providing affordable and practical access to the detectors' raw output, we can drive the development of efficient algorithms that fully exploit the advantages of fast direct electron detectors.

# Results

## Data Reduction Levels

A secondary electron puddle is characterized by its spatial location, its two-dimensional (2D) shape, and the pixel intensities within this shape. To accommodate various downstream processing needs we define four logical data reduction levels: L1, L2, L3, and L4, with progressively higher levels of file size reduction and hence information loss from L1 to L4 (Figure 1).

**<Figure 1: Data reduction levels>**

All four data reduction schemes begin with a thresholding step, which produces a binary map identifying pixels as containing useful signal or not. The ADU (analog-digital unit) threshold used to label signal pixels are independent for each pixel and decided based on the signal-noise calibration procedure (discussed below). In

ReCoDe level L1, the sparsified signal pixel intensities are then bit-packed into a dense format. Bit packing removes unused bits and converts the list of ADU values into a continuous string of bits. The binary map and the bit packed intensity values are independently compressed and the two compressed data are stacked to create an L1 reduced compressed frame. As L1 reduction retains all the information about electron puddles, electron counting can be performed long afterward, should the user wish to explore different counting algorithms and/or parameters. Both thresholding and packing are sufficiently fast to make L1 suitable for on-the-fly processing (discussed in "Demonstration of on-the-fly Reduction and Compression" section). Even for relatively high electron flux data (0.05 e/pixel/frame) L1 reduction alone achieves a 10x reduction in file size. This reduced data can be further compressed to achieve an overall 25x file size reduction (Figure 3).

In L3 reduction, the pixel intensities are discarded during thresholding and only the binary map is retained and compressed. L3 is therefore optimized for speed, at the expense of puddle specific ADU (pixel intensity) information.

To compute puddle specific features, in L2 and L4 reductions, the clusters of connected pixels that constitute individual puddles are identified from the binary map using a connected components labeling algorithm, discussed in the Methods section. In L4 reduction, each puddle in the binary map is further reduced to the single pixel, where the primary electron was likely incident. L4 reduction, therefore, results in a highly sparse binary map that is optimized for maximum compression. At the same electron flux (0.05 e/pixel/frame) L4 reduction and compression results in 45x file size reduction. This increased compression comes at the cost of throughput since counting has to be performed as part of the reduction step.

In L2 reduction, a summary statistic, such as mean, maximum or sum of ADU, is extracted for each electron puddle. Preliminary studies suggest that such information may correlate with whether a measured electron was elastically or inelastically scattered[1]. The sparse puddle features are then packed into a dense format and the binary map and the dense puddle features are independently compressed. Several applications that record diffraction patterns benefit from a high dynamic range but do not necessarily need to retain the entire signal as done in L1. L2 is designed for such applications.

In L1 and L2 reductions, the binary maps and the packed intensity summary statistics are independently compressed and then stacked. As the binary maps and intensity values have very different characteristics, compressing them independently results in optimal compression (Fig. 1b-d).

The reduced compressed data formats are detailed in Supplementary Method S1.

All four data reduction schemes in ReCoDe first reduce the data by removing primarily readout noise (thresholding) and then compressing the signal. Accurate signal-noise separation is therefore critical. To remove pixel-level differences in dark noise and gain that can bias the identification of isolated electron puddles, individual thresholds are calculated per pixel based on calibration data (Methods section). For the DE-16 detector, this calibration can be done with a single dataset with flat-field illumination at a low dose rate and extended exposure times. Since different detectors may require custom calibration, ReCoDe only requires the per pixel thresholds for separating signal-noise as input and is agnostic of the calibration method used. These thresholds are specified in a single image frame, which is reloaded by ReCoDe at user-specified intervals. External programs can update the thresholds intermittently for on-the-fly recalibration to accommodate changing detector response.

## Calibrating parameters for data reduction

An appropriate threshold separating signal from noise is critical for electron counting to be effective. Typically, this threshold is established through calibration, based on dark and gain references obtained during data acquisition. In most imaging software these calibrations depend on several hyper-parameters that are predetermined (for instance, the number of frames used in the dark reference). Once the calibrated frames are reduced to electron counted images, the calibration cannot be revised, and the effects of the hyper-parameters are permanent. The L1 reduction presents an alternative, where the data can be recalibrated post-acquisition without having to store the entire dataset, as long as a sufficiently permissive threshold is used. In low dose rate experiments, during data acquisition, the quality of images cannot be verified through visual inspection. The effectiveness of the calibration can, therefore, be difficult to judge. The ability to recalibrate datasets in such cases can significantly improve image quality, as shown in Figure 2. Here, the data was recalibrated by using a higher threshold for separating dark noise and signal and pixel gains were recalculated after removing single pixel puddles (see Fine calibration in Methods Section for details). We observed that such recalibration can significantly reduce the number of false positive electron events.

**<Figure 2: L1 Recalibration of MoS2 sample>**

Even small deviations in calibration can significantly bias counting and therefore recalibration (or at least a quality assessment) should be a necessary step in ensuring accurate counting. L1 reduced data facilitates such post-hoc analysis. This includes using the electron puddle size/shape distributions to estimate realistic coincidence losses specific to the detector and imaging conditions (Table 1).

**<Table 1: Coincidence loss estimation methods.>**

Table 1 shows coincidence losses estimated using five different techniques. In the first three (columns from left to right) puddles are assumed to be of fixed shape and size, whereas, in the last two, the actual puddle shape and size information are included in the calculation (see Methods section). Clearly, the knowledge of puddle shape and size is essential for accurate coincidence loss estimation. Therefore, accurately estimating coincidence loss requires retaining data at reduction levels L1-L3.

A recent study[8] has proposed storing L4 reduced data in a sparse format to benefit from higher dose fractionation without overwhelming acquisition systems with storage requirements. To achieve super-resolution electron counting, which is critical for improving reconstruction resolution in cryo-EM, they propose subdividing each pixel before counting and storing the higher-resolution spatial locations of electron events using a higher bit-depth. ReCoDe's L1 reduction scheme enables super-resolution electron counting without the need to subdivide pixels at the time of acquisition, thus eliminating the need to predetermine to what extent pixels should be partitioned.

## Reducibility and Compressibility with Increasing Electron Fluxes

With increasing electron flux, the data naturally becomes less reducible and less compressible. To quantify this change, we simulated images at eight electron fluxes between 0.0025 to 0.07 e/pixel/frame (Figure 3). This range was chosen for tolerable coincidence loss during electron counting (Table 1). For data without any reduction (unreduced compression line in Figure 3), the compression ratio remains similar across all fluxes (~4x), because of readout dark noise. L3 and L4 reduced data are essentially binary images with 1-bit per pixel (Figure 1). Therefore, if the input data uses $n$ bits to represent each pixel's intensity, a factor of $n$ reduction is achieved using L3 or L4 reduction alone. In Figure 3, a 16x reduction is seen for the 16-bit simulated data. In L1 and L2, pixel intensity information and event summary statistics are retained in addition to the L3 binary map. As electron flux increases, more pixel intensities/event statistics need to be stored. However, due to coincidence loss the number of counted electron events and L1 and L2 file sizes increase only sub-linearly.

With increasing electron flux the binary images used to store location and shape information in the reduced format, also become less compressible. This is evident from the L3 and L4 "reduction + compression" lines in Figure 3. At the same time, for L1 reduction, the proportion of reduced data containing pixel intensities increases rapidly with increasing electron flux. As a result, the compressibility of L1 reduced data falls very quickly with increasing electron flux.

At moderate (0.01 e/pixel/frame) and low (0.001 e/pixel/frame) electron flux L1 reduction compression results in 60x and 170x data reduction, respectively.

Reduction L4, where only puddle locations are retained, is optimized for maximum compression and can achieve reduction compression ratios as high as 45x, 100x and 250x at high (0.05 e/pixel/frame), moderate (0.01 e/pixel/frame) and low (0.001 e/pixel/frame) electron flux, respectively.

Compression algorithms exploit the same basic idea: by representing the more frequently occurring symbols with fewer bits the total number of bits needed to encode a dataset is effectively reduced. Consequently, data is more compressible when symbols are sparsely distributed. Such sparse distributions are readily present in the back-thinned DE-16 electron detector, where nearly 80% of the digitized secondary electron puddles span fewer than three pixels (Supplementary Fig. S8). Even for puddles that span four pixels (of which there are 110 possibilities) nearly half (48.3%) are the 2x2-pixel square motif.

The randomly distributed centroids of secondary electron puddles account for the largest fraction of memory needed to store the reduced frames. We considered three representations for storing these centroids and ultimately adopted a binary image representation (Methods section).

<Figure 3: Reducibility and compressibility of data with increasing electron flux.>

## Compression Algorithms

Any lossless compression algorithm can operate on the reduced data levels in Figure 1. Compression algorithms are either optimized for compression power or for compression speed, and the desired balance depends on the application. For on-the-fly compression, a faster algorithm is preferable even if it sub-optimally compresses the data, whereas an archival application may prefer higher compression power at the expense of compression speed.

We evaluated the compression powers and speeds of six popular compression algorithms that are included by default in the ReCoDe package: Deflate[15], Zstandard (Zstd), bzip2 (Bzip)[16,17], LZMA[18], LZ4[19] and Snappy[20] (Figure 4). Each algorithm offers different advantages; bzip, for instance, is optimized for compression power whereas Snappy is optimized for compression and decompression speed. All five algorithms can be further parameterized to favor compression speed or power. We evaluated the two extreme internal optimization levels of these algorithms: fastest but sub-optimal compression, and slowest but optimal compression.

Data reduction schemes similar to L1 have been previously used to compress astronomical radio data in Masui et al.[21]. They proposed the bitshuffle algorithm for compressing radio data after removing thermal noise from it. We experimented with Blosc, a meta-compressor for binary data, that implements bitshuffle in addition to breaking the data into chunks that fit into the system's L1 cache, to improve compression throughputs (Figure 4).

<Figure 4: Comparison of compression algorithms for different reduction levels and dose rates.>

LZ4 and SNAPPY have the highest throughputs across all reduction levels and electron fluxes, with reduction compression ratios slightly worse than the remaining five algorithms. At the lowest dose rate (0.01 e/pixel/frame) Bzip results in the best reduction compression ratios, regardless of the internal optimization level. At higher dose rates (0.03 and 0.05 e/pixel/frame) Zstd has the highest compression ratio. Considering all dose rates and internal optimization levels, Zstd on average offers the best balance between compression ratio and throughput. The choice of internal optimization level only marginally affects the reduction compression level but significantly improves throughput.

Deflate optimized for speed for instance is almost ~25x faster than Deflate optimized for compression, across the three dose rates. We use Deflate optimized for speed (referred to as Deflate-1) as the reference compression algorithm for the rest of the paper, as it represents a good average case performance among all the compression algorithms. In subsequent sections, we will show that Deflate-1 is fast enough for on-the-fly compression. All algorithms have higher decompression throughput than compression throughput (Supplementary Fig. S2). Deflate-1 has ten times higher decompression throughput than compression throughput, which means the same computing hardware for reduction and compression can support on-the-fly retrieval and decompression of frames for downstream data processing.

For most compression algorithms Blosc marginally improves compression throughput (Figure 4), except in the case of optimal compression with LZ4, where Blosc improves throughput by as much as 400 MB/s.

## Demonstration of on-the-fly Reduction and Compression

Electron microscopy imaging often has to be performed at low electron flux to reduce beam induced damage, if the sample is dose sensitive, as well as to minimize beam induced reactions. Observing rare events or slow reactions in such cases require extended acquisition, that is not feasible with current detector software without compromising temporal resolution. Loss of temporal resolution, in turn, degrades drift correction and therefore limits spatial resolution. ReCoDe's on-the-fly reduction compression fills this critical gap, enabling hours long continuous acquisition without overwhelming storage requirements, compromising on temporal resolution, or losing puddle information.

ReCoDe is easily parallelized, with multiple threads independently reducing and compressing different frames in a frame stack. In this multithreaded scheme, each thread reduces and compresses the data to an intermediate file, which are merged when data collection is complete. The merging algorithm reads from the intermediate files and writes to the merged file sequentially and is therefore extremely fast. The intermediate and merged (ReCoDe) file structures and the merging process are described in Supplementary Method S1.

With this multithreaded scheme, ReCoDe can achieve throughputs matching that of the detectors enabling on-the-fly reduction and compression. Additionally, intermediate files can be accessed sequentially in both forward and reverse directions, with frames indexed by frame number, time stamp, and optionally scan position. Owing to the small size of the reduced compressed frames, they can be read from intermediate files by external programs for live processing and feedback during acquisition even without merging them back into a single file. Users also have the option of retaining raw (unreduced and uncompressed) frames at specified intervals for validation or for on-the-fly recalibration. In electron microscopy facilities data is often archived in high capacity network-attached storage (NAS) servers. A schematic of this on-the-fly reduction compression pipeline is shown in Figure 5a. We evaluated the feasibility of directly collecting the reduced-compressed data onto NAS servers, to avoid the overhead of transferring data after collecting it on the microscope's local computer.

With the DE-16 detector running at 400 fps, at a dose rate of 0.001 e/pixel/frame and ReCoDe using 10 CPU cores of the acquisition computer that shipped with the DE-16 detector, we continuously captured data directly onto NAS servers connected by a 10 gigabits/s Ethernet (10 GbE) connection, for 90 minutes (Methods subsection: *On-the-fly Compression Pipeline*). To further evaluate this multithreaded scheme, we simulated a series of on-the-fly data reduction and compression at different electron fluxes. The implementation used for these simulations emulates the worst-case write performance of ReCoDe, where a single thread sequentially accesses the disk (see Supplementary Discussion S3 for details). At relatively low electron flux (0.01 e/pixel/s) we are able to achieve throughputs as high as 8.3 gigabytes per second (GB/s, Figure 5b) using 50 threads on a 28 core system. At the same dose rate, to keep up with the DE-16 detector (which has a throughput of ~3.08 GB/s) only 10 CPU cores are sufficient. For perspective, another popular direct electron detector, the K2-IS (Gatan Inc.), nominally outputs bit-packed binary files at approximately 2.2 GB/s. However, since we did not have to incur extra computation time to unpack bits on the raw data from DE-16, the DE-16 benchmarks on Figure 5 will not directly apply to K2-IS data.

**<Figure 5: On-the-fly reduction compression>**

At moderate electron flux, writing directly to GPFS NAS servers using both 10 GbE and IPoIB (Internet Protocol over InfiniBand) has comparable throughputs to that of collecting data locally on the microscope's computer (Figures 5c and 5d). However, at very low electron flux writing directly to the NAS server with IPoIB has slightly higher throughput. This is likely due to the reduced communication overhead per call in IPoIB and the distributed data access (IBM GPFS) supported by NAS servers, both of which are optimized to handle multiple simultaneous small write requests. In the absence of such a parallel data access ReCoDe still executes at close to 89% parallel (Supplementary Discussion S3).

Both the reduction and compression steps are essential for high throughput on-the-fly processing. Without compression, the reduced data is still too large to write over 10GbE, particularly at moderate electron flux (Figure 5e). Without reduction, the data is not compressible enough; the throughput of Deflate-1 compression without any data reduction (Figure 5f) is abysmally low even when using 50 threads.

## Discussion

Studying millisecond *in-situ* dynamics with TEM, such as surface-mediated nanoparticle diffusion in water[22], requires us to operate at the maximum frame rates of these detectors. Additionally, longer total acquisition times would be beneficial for studying reactions such as spontaneous nucleation[23] where the experimenter systematically searches a large surface for samples. Several pixelated TEM electron detectors are now able to achieve sub-millisecond temporal resolutions, with the downside that the local buffer storage accessible to these detectors fills up very quickly. Figure 7a shows that current TEM detectors running at maximum frame rates produce 1 TB of data in several minutes. When the temporal resolution is critical for an imaging modality, reducing the frame rate is not an option. An example is fast *operando* electron tomography[24]. To capture how the 3D morphology of an object evolves over several seconds, a full-tilt series of the object has to be rapidly acquired at the detector's peak frame rate. Here again, the duration of these observations can be significantly extended by substantially reducing the output data load with ReCoDe.

**<Figure 6: The storage costs of TEM experiments with movie-mode detectors. >**

4D scanning transmission electron microscopy (4D STEM) techniques including Ptychography were used to image weak phase objects and beam sensitive samples such as Metal Oxide Frameworks (MOFs)[25]. Here, a converged electron probe raster scans a sample collecting 2D diffraction patterns at each scan point. Although these experiments can produce hundreds of gigabytes of data in minutes[26], the diffraction patterns tend to be sparse outside of the central diffraction spot. As noise-robust STEM-Ptychography becomes a reality[27], their convergent beam electron diffraction patterns will be even sparser. ReCoDe level L1 reduction and compression, which preserves the patterns' dynamic range while removing only dark noise, are likely to be useful for such data. Once the large datasets in 4D STEM are reduced they will readily fit into the RAM of desktop workstations, which also facilitates sparse and efficient implementations of processing algorithms.

Electron beam-induced damage is a major limitation for all cryo-EM modalities. In single-particle analysis (SPA) the energy deposited by inelastically scattered electrons manifests as sample damage and ice drift, where global and site-specific sample damage is detectable even at exposures as low as 0.1 e/$Å^2$ [28]. Here higher electron dose fractionation improves resolution in two ways: (1) by reducing coincidence loss and thereby improving detection efficiency[29] and (2) by enabling more accurate estimation of sample drift at a higher temporal resolution [8]. Increasing detector frame rates can reduce the average displacement of each particle captured in each dose-fractionated frame, but doing so further inflates the already large amounts of movie-mode data collected (see Supplementary Fig. S6). On-the-fly reduction and compression can significantly reduce the storage costs of movie-mode data, to accommodate image correction algorithms that operate at a degree of dose fractionation that is higher than current practice.

The recently proposed compressed MRCZ format[30] and ReCoDe offer complementary strategies to reduce file sizes generated by electron detectors. MRCZ is ideal for compressing information-dense images of electron counts integrated over longer acquisition times. ReCoDe, however, excels in reducing and compressing the much sparser raw detector data that are used to produce the integrated images typically meant for MRCZ. By doing so

ReCoDe can preserve the arrival times of incident electrons that are lost when they are integrated into a single frame. Applying an MRCZ-like scheme on the raw un-reduced signal is inefficient, as shown with the "Unreduced Compression" line in Figure 3. Figure 7a compares compression ratios obtained by MRCZ and ReCoDe on publicly available EMPIAR datasets from multiple published results[31–34], spanning a range of dose rates and detectors as listed in Figure 7c. Figure 7b shows the compression ratios achieved by the two approaches on simulated images. Across the range of dose rates ReCoDe produces better compression ratios on the EMPIAR datasets. As low dose rate datasets (below 0.58 electrons/pixel/frame) are likely to be sparse, ReCoDe as expected, achieves higher compression ratios than MRCZ (Figure 7b). However, surprisingly, ReCoDe outperforms MRCZ even for some datasets with much higher average dose rates (EMPIAR-10346 in Figure 7a). These datasets have particularly high contrast resulting in higher average dose rates but are still very sparse (Supplementary Fig. S7c), making them well suited for compression with ReCoDe.

**<Figure 7: Comparison of ReCoDe and MRCZ.>**

We have described three novel analysis methods that demonstrate the necessity of reduction levels L1-L3. These methods cannot be applied on counted (L4 reduced) data, as they rely on puddle shape and intensity information. The first is the recalibration of L1 reduced data post acquisition, to improve counting accuracy. The second analysis uses puddle shape information to accurately estimate coincidence loss. When counting electrons, coincidence loss adversely affects spatial resolution (Supplementary Figure S5). However, as we have shown, estimates of coincidence loss from the counted data can be inaccurate (Table 1). As reduction levels L1-L3 retain puddle shape information these can be used when accurate coincidence loss estimates are desired. In the third analysis we use a series of L1 reduced data sets with diminishing dose rates and extremely sparse electron events to estimate false positive rates of detecting electron events (Supplementary Note S12).

We also describe a novel method for estimating the proportion of backscattered electrons, using counted (L4 reduced) data (Supplementary Note S13). Using this analysis we estimated the proportion of primary to backscattered electrons for the DE-16 detector is ~8.6. In the future, it may be possible to even classify and eliminate backscattered electrons based on their sizes, shapes and proximity to primary electrons. Development of such techniques requires retaining more information than is currently done using counted data. The L1-L3 reduction levels in ReCoDe are designed to facilitate such future developments.

In summary, we present the ReCoDe data reduction and compression framework for high-throughput electron-counting detectors, which comprises interchangeable components that can be easily configured to meet application-specific requirements. ReCoDe supports four data reduction levels to balance application-specific needs for information preservation and processing speed. We further tested three electron localization strategies, and show that they produce similar spatial resolutions even when the electron puddle intensity information is absent. By comparing five candidate compression algorithms on reduced electron data, we found that although LZ4 is the fastest, Deflate-1 offers the best compromise between speed and compressibility

Remarkably, we demonstrated on-the-fly data reduction and compression with ReCoDe on the DE-16 detector for 90 minutes. Using only a desktop workstation, we continuously converted a 3 GB/s raw input data stream into a ~200 MB/s output that was, in turn, streamed onto networked drives via 10 Gbit ethernet. Crucially, this demonstration showed that on-the-fly data reduction and compression at low dose rates on our fastest S/TEM detectors is not compute-limited if the detector's raw data stream is accessible (via a RAM-disk) before it is stored to SSDs. Even higher throughputs will be achievable with direct in-memory access to this raw data stream without the need for a RAM-disk.

The ReCoDe scheme can dramatically increase the throughput of electron microscopy experiments. Furthermore, the quality of observations for electron microscopy experiments can also improve. In cryo-EM, ReCoDe can support movies of higher frame rates, which can lead to better drift correction and lower coincidence loss. For *in-situ* experiments, higher frame rates can also improve the temporal and spatial resolution of the imaged samples.

Currently, a clear barrier for commercial vendors to produce higher throughput detectors is that users cannot afford to store the increased raw data that these faster detectors will bring. By efficiently reducing raw data into compact representations, ReCoDe prepares us for an exciting future of megahertz electron detectors in three crucial ways: it limits the storage costs of electron microscopy experiments, facilitates much longer data acquisition experimental runs, and very efficient processing algorithms that only compute on the essential features. More broadly, making ReCoDe open source encourages its own development by the community and incentivizes commercial vendors to specialize in much-needed hardware innovation. The full impact of electron counting detectors, quite possibly, is still ahead of us.

## Acknowledgments

The authors would like to thank Xiaoxu Zhao for his kind contribution of the molybdenum disulphide 2d crystals. We also thank Liang Jin for helpful discussions, and Benjamin Bammes regarding the DE-16 detector and data processing pipeline, plus his careful reading of the manuscript. We are grateful to Ming Pan, Ana Pakzad, and Cory Czarnik for details about the K2-IS detector and associated binary data formats. The authors would also like to acknowledge Chong Ping Lee from the Centre for Bio-Imaging Sciences (CBIS) at the National University of Singapore (NUS) for training and microscope facility management support, and Bai Chang from CBIS for IT infrastructure support. A.D. and Y.B. were funded by the Singapore National Research Foundation (grant NRF-CRP16-2015-05), with additional support from the NUS Startup grant (R-154-000-A09-133), NUS Early Career Research Award (R-154-000-B35-133), and the Singapore Ministry of Education Academic Research Fund Tier 1 Grant (R-154-000-C01-114).

## Author contributions

D.L. and A.D. conceived the project. A.D., D.B., K.F.N., S.W.C., and J.S. collected data, which A.D., K.F.N., and M.D. analyzed. ReCoDe framework was programmed by A.D., under the D.L.'s advice. A.D. and D.L. wrote the manuscript with inputs from all the authors.

## Competing interests

The authors declare that they have no conflict of interests.

## Methods

### Data Acquisition

All experimental data were collected on a DE-16 detector (Direct Electron Inc., USA) installed on the JEM-2200FS microscope (JEOL Inc., Tokyo, Japan) equipped with a field emission gun and operating at 200 keV accelerating voltage. StreamPix (Norpix Inc., Montreal, Canada) acquisition software was used to save the data in sequence file format without any internal compression. Data for puddle shape and size analysis (Supplementary Fig. S8) and MTF characterization with the knife-edge method (Figure 5) were collected at 690 frames per second and 400 frames per second respectively with an electron flux of ~0.8 e/pixel/s.

All simulations of on-the-fly data collection were performed on a 28-core (14 core x 2 chips) system with 2.6 GHz E5-2690v4 Intel Broadwell Xeon processors and 512 GB DDR4-2400 RAM.

### Connected Components Labelling

To compute the features specific to each electron puddle (e.g. centroids in L4 and the user-chosen summary statistics (ADU sum or maximum) in L2), the set of connected pixels (components) that constitute individual puddles have to be identified from the thresholded image. This connected components labeling can be computationally expensive for large puddles. Fortunately, puddle sizes tend to be small for most back-thinned direct electron detectors. For the DE-16 detector, 90% of the puddles are fewer than five pixels in size (Supplementary Fig. S8). Therefore, we use a greedy approach similar to the watershed segmentation algorithm[35] to perform connected components labeling. The algorithm assigns unique labels to each connected component or puddle, and the pixels associated with a given puddle are identified by the label of that puddle. In L2 reduction, these labels are used to extract the chosen summary statistics from the puddle and in L4 reduction, these labels are used to approximate the secondary electron puddle to a single pixel, by computing the centroid or center of mass, etc.).

### Representation of Puddle Centroids

The randomly distributed centroids of secondary electron puddles account for the largest fraction of memory needed to store the reduced frames. We considered three representations for storing these centroids In the first representation, a centroid is encoded as a single 2n-bit linear index. In the second representation, these linear indices are sorted and run-length encoded (RLE) since the ordering of centroids in a single frame is inconsequential. In the third representation, the centroids are encoded as a binary image (similar to L4). The RLE and binary image representations were found to be much more compressible than linear indices (Supplementary Fig. S9). Ultimately, the binary image representation was adopted in ReCoDe because the sorting needed for RLE is computationally expensive, for only a marginally higher compression.

### Signal-Noise Calibration

In the current implementation, ReCoDe requires as input a single set of pre-computed calibration parameters, comprising each pixel's dark and gain corrected threshold for separating signal and noise at that pixel. Any calibration method can be used to compute this calibration frame. The *On-the-fly Calibration* Methods subsection below describes a fast routine, for estimating this calibration frame, which we applied to the DE-16 detector. The "Fine Calibration" Methods subsection thereafter details a more deliberate data collection approach, where additional diagnostics on the detector are also measured. Both calibration approaches yield practically similar results at dose rates above $10^{-4}$ e/pix/frame (Supplementary Note S14).

### On-the-fly Calibration

First, a flat-field illuminated dataset, comprising many raw detector frames preferably at the same low dose rate targeted for actual imaging afterward, is collected. An estimate of the incident dose rate is then computed using this dataset. Whereas this could be obtained with an independent measurement (e.g. Faraday cup), the dose rate computed by the procedure described here factors in the detector's detective quantum efficiency. Ideally, a pixel's intensity across the calibration frames would follow a mixture of two well-separated normal distributions, corresponding to either dark noise or signal. However, in practice, because the detector PSF is larger than a pixel, charge sharing from fast electrons incident on neighboring pixels will contribute to a single pixel's intensity, which causes the noise and signal distributions to overlap severely (Supplementary Fig. S10).

The calibration (summarized in Supplementary Note S11) begins by first estimating a single global threshold that separates signal from noise for all pixels. Assuming the histogram of dark values are normally distributed, this global threshold is estimated based on a user-specified upper limit on the tolerable false positive rate of a surmised normal distribution $(r)$. However, because individual pixels behave differently from each other, using the same threshold for all pixels can severely bias electron counting. To remove this bias the global threshold has to be adapted for each pixel individually based on the pixel's gain and dark noise level.

Now we are ready to estimate the effective detectable electron count on the detector from the dataset directly. Given the low dose rate in this calibration dataset, only in a small fraction of frames does an individual pixel see electron events. Therefore, a pixel's median across all calibration frames is effectively its dark noise level, at this dose rate. Given the compact PSF and high SNR of DE-16 detectors, to calculate each pixel's gain, we assume that direct electron hits result in larger intensities than those due to charge sharing, even when the pixels have different gains. If the calibration dataset has a total dose of $N$ e/pixel, where $N$ is sufficiently small such that the probability of two electrons hitting the same pixel is negligible, then a pixel's gain is the median of the $N$ largest intensities it has across all calibration frames. Therefore, we first estimate the total dose per pixel in the calibration dataset using a few randomly selected small two-dimensional (2D) patches. Separate thresholds are identified for individual pixels in these patches in a similar manner to the global threshold (i.e. assuming normality in the dark distribution and using a false positive rate parameter $r$). These thresholds are used to identify the connected components in each selected 2D patch across all frames in the calibration dataset. The number of connected components emanating from the central pixel of a 2D patch across all calibration frames gives an estimate of the number of electron events ($n_c$) at the central pixel of that patch. The average of these values across all randomly selected patches ($\bar{n}_c$) is used as the estimated total dose per pixel in the calibration dataset. Here, a puddle is assumed to emanate from the pixel that has the maximum value in the puddle. Finally, using the per-pixel dark noise levels and gains the global threshold is adapted to compute each pixel's independent threshold. To compute a pixel's threshold the global threshold is first shifted such that the pixel's dark noise level matches the global mean dark noise level and then scaled such that the pixel's gain matches the global gain.

For sufficiently sparse calibration data, even mean pixel intensity, which is much more efficiently computed than median, can be used to estimate the dark noise level for the pixel, although at the expense of a slightly higher false positive rate.

## Fine Calibration

To further assess the fast, on-the-fly calibration, a slower and more intricate calibration, referred to as *fine calibration*, was also implemented. The fine calibration adds two steps to the on-the-fly calibration procedure described above, a common-mode correction and a puddle area based filtering. The common-mode correction eliminates dynamically fluctuating biases in electron counting due to correlated thermal fluctuations between pixels that are connected to the same local voltage (hence thermal) source; the area threshold filters electron puddles to reduce false positive puddle detection. Analysis of the temporal response of DE-16's pixels revealed local detector regions of size 4x256 pixels that have correlated responses (Supplementary Note S12 and Supplementary Figure S12). The fine calibration steps are summarised in Supplementary Figure S13.

A series of datasets with increasingly sparse data was used to compare the two calibrations. The controlled reduction in dose rate was achieved by increasing magnification in successive datasets by 200x while keeping the electron flux constant. A comparison of the rate of decay of the estimated dose rates with counting following the two calibration strategies revealed that on-the-fly calibration includes a substantial number of false positive puddles (Supplementary Figure S14A-B). However, this can be easily remedied with area-based filtering following L1 reduced data acquisition. While the common-mode correction has only a marginal effect on counting with the DE-16 detector (Supplementary S14C), it might be essential for other detectors.

Backthinning of direct electron detectors was instrumental in reducing noise from backscattered electrons while also shrinking electron puddle sizes[36]. The smaller puddle sizes and improved signal-to-noise ratio, in turn, made electron counting feasible. By comparing the distribution of neighboring puddle distances in the ultra-low dose rate datasets with those in simulated images, we were able to estimate the ratio of primary to backscattered electrons to be ~10.1 (Supplementary Note S15). A feature of L1 reduction is that all the puddle shape information is retained. In the future this shape information may be useful in algorithmically distinguishing backscattered electrons from primary electrons, leading to a further reduction in noise due to backscattered electrons.

## On-the-fly Compression Pipeline

Continuous on-the-fly data reduction and compression for 90 minutes were performed using 10 cores of the computer shipped with the DE-16 detector. This computer has two E5-2687v4 Intel Xeon processors (24-cores, 12 cores per chip, each core running at an average of 3.0GHz base clock rate), 128 GB DDR4 RAM, and is connected to a 1 Petabyte IBM GPFS NAS via a 10 GbE connection.

If the raw data stream coming from the detector is accessible in-memory (RAM), reduction compression can be performed directly on the incoming data stream. However, many detectors (including the DE-16 detector) make the raw data available only after it is written to disk. Reduction compression then requires simultaneously reading the data from disk back into RAM while more data from the detector is being written to disk. While sufficiently fast SSDs in RAID 0 can support multiplexed reads and writes to different parts of the RAID partition, a more scalable solution is to use a virtual file pointer to a location in fast DDR RAM (using RAM-disk). DDR RAMs, in fact, have sufficient read-write throughputs such that multiple ReCoDe threads can read different sections of the available data stream parallelly, while new data coming from the detector is written.

For continuous on-the-fly data collection, the StreamPix software was configured with a script to acquire data in five-second chunks and save each chunk in a separate file. The DE-16 acquisition software does not allow direct in-memory access to data coming from the detector to RAM, restricting access to data only after it has been written to disk. While SSDs have fast enough write speeds to keep up with the throughput of the DE-16 detector, on-the-fly reduction compression requires simultaneous write and read, each at 3GB/s, which is not possible even with SSDs. To overcome this problem, a virtual file pointer to a location in fast DDR RAM (using RAM-disk) was used. When StreamPix finishes writing a five-second file to RAM-disk, the ReCoDe queue manager adds the file to the queue and informs the ReCoDe server. The ReCoDe server then picks off the next five-second file in the queue, where each processing thread in the server independently reads a different subset of frames within this file. Subsequently, each thread independently appends its reduced and compressed output to its own intermediate file on the NAS server via the 10 GigE connection. When the ReCoDe server is finished processing a five-second file in the queue it informs ReCoDe queue manager, which then deletes this file from RAM-disk. When the acquisition is complete all intermediate files are automatically merged into a single ReCoDe file where the reduced and compressed frames are time-ordered.

While the RAM-disk based approach bypasses read-writes to SSDs, it requires copying the same data in RAM twice. First, the data stream from the detector is written to an inaccessible partition on the RAM, then copied to the readable RAM-disk partition. If we had direct access to first copy in the currently inaccessible partition on the RAM, the subsequent copy to the RAM-disk can be eliminated, hence freeing up important read-write bandwidth on the RAM. At the DE-16 detector's throughput of 3.08 GB/s, this copying (read and write) uses a significant portion of the RAM's bandwidth (6.16 GB/s out of DDR4 RAM's 21-27 GB/s, or 20-25 GiB/s, transfer rate). Direct access to the detector's data stream without such copying will, therefore, enable reduction compression at even higher throughputs.

A fully parallelized Pythonic implementation of ReCoDe with features for on-the-fly reduction compression is available at https://github.com/NDLOHGRP/pyReCoDe.

## Effects of Reduction and Coincidence Loss on Counted Image Quality

In many applications, the L2 and L3 reduced data has to be ultimately reduced to L4 (electron-counted image). Here, we consider how the information lost in L2 and L3 reductions affect the resolution in L4 images. In L4 puddles are reduced to a single pixel, which ideally contains the entry point of the incident electron. However, there is no clear consensus on the best approximation strategy for determining an electron's entry point, given a secondary electron puddle[1]. The three common strategies are, to reduce the puddle to 1) the pixel that has the maximum intensity, 2) the pixel intensity weighted centroid (center of mass) or 3) the unweighted centroid of the puddle. Unlike L1 reduction, where all the information needed for counting with any of these strategies are retained, with L2 and L3 reductions, the pixel intensity information is either partially or completely lost. The puddles can then only be reduced to the unweighted centroid of the puddle using the third strategy. With L4 reduction, the approximation strategy has to be chosen prior to data acquisition. To evaluate how this information loss affects image quality we performed a knife-edge test using a beam blanker (see Methods section for implementation details). The results show (Supplementary Fig. S4) that the choice of approximation strategy, and therefore the choice of reduction level, has little consequence on image resolution.

Coincidence loss not only reduces the amount of signal transferred onto the electron-counted image, reducing $DQE(0)$ [36,37], it also increases the localization error when reducing puddles to a single pixel. When puddles due to separate incident electrons merge, the probability that the approximation strategy incorrectly identifies the entry point of the primary electron increases. MTFs corresponding to counted images at electron fluxes ranging from 0.005 to 0.1 $e^-/Å^2/s$ show that increasing electron flux reduces MTF at higher frequencies (Supplementary Fig. S5).

To accurately estimate the effect of coincidence loss on counting, we simulated images where puddles follow the shape and size distributions of those in real DE-16 data acquired at 0.001 $e^-/pixel/frame$. Table 1 lists coincidence losses estimated from the simulated images following five different approaches. In the first three approaches the puddles are assumed to be of fixed shape and size, in the fourth approach the puddles vary in size but are symmetric in shape, while in the fifth approach puddles follow shape and size distributions estimated from actual DE-16 data (see Supplementary Discussion S5). Clearly, accurate coincidence loss calculation requires factoring in the shape and size of secondary electron puddles, without which, the results are markedly different.

## References

1. Datta, A., Chee, S. W., Bammes, B., Jin, L. & Loh, D. What Can We Learn from the Shapes of Secondary Electron Puddles on Direct Electron Detectors? *Microsc. Microanal.* **23**, 190–191 (2017).

2. Li, X., Zheng, S. Q., Egami, K., Agard, D. A. & Cheng, Y. Influence of electron dose rate on electron counting images recorded with the K2 camera. *J. Struct. Biol.* **184**, 251–260 (2013).

3. Johnson, I. J. *et al.* Development of a fast framing detector for electron microscopy. in *2016 IEEE Nuclear Science Symposium, Medical Imaging Conference and Room-Temperature Semiconductor Detector Workshop (NSS/MIC/RTSD)* 1–2 (2016).

4. Chee, S. W., Anand, U., Bisht, G., Tan, S. F. & Mirsaidov, U. Direct Observations of the Rotation and Translation of Anisotropic Nanoparticles Adsorbed at a Liquid-Solid Interface. *Nano Lett.* **19**, 2871–2878 (2019).

5. Levin, B. D. A., Lawrence, E. L. & Crozier, P. A. Tracking the picoscale spatial motion of atomic columns during dynamic structural change. *Ultramicroscopy* **213**, 112978 (2020).

6. Liao, H.-G. *et al.* Nanoparticle growth. Facet development during platinum nanocube growth. *Science* **345**, 916–919 (2014).

7. Zheng, S. Q. *et al.* MotionCor2: anisotropic correction of beam-induced motion for improved cryo-electron microscopy. *Nat. Methods* **14**, 331–332 (2017).

8. Guo, H., Franken, E., Deng, Y., Benlekbir, S. & Lezcano, G. S. Electron Event Representation (EER) data enables efficient cryoEM file storage with full preservation of spatial and temporal resolution. *bioRxiv* (2020).

9. Allahgholi, A. *et al.* AGIPD, a high dynamic range fast detector for the European XFEL. *J. Instrum.* **10**, C01023 (2015).

10. El-Desouki, M. *et al.* CMOS Image Sensors for High Speed Applications. *Sensors* **9**, 430–444 (2009).

11. Iudin, A., Korir, P. K., Salavert-Torres, J., Kleywegt, G. J. & Patwardhan, A. EMPIAR: a public archive for raw electron microscopy image data. *Nat. Methods* **13**, 387–388 (2016).

12. Henderson, R. *et al.* Outcome of the first electron microscopy validation task force meeting. *Structure* **20**, 205–214 (2012).

13. McCallum, J. C. Disk Drive Prices (1955-2019). https://jcmit.net/diskprice.htm.

14. Klein, A. The Cost of Hard Drives Over Time. *Backblaze Blog | Cloud Storage & Cloud Backup* https://www.backblaze.com/blog/hard-drive-cost-per-gigabyte/ (2017).

15. Deutsch, L. P. DEFLATE compressed data format specification version 1.3. (1996).

16. M. Burrows, D. J. W. A block-sorting lossless data compression algorithm. (1994).

17. bzip2 : Home. https://www.sourceware.org/bzip2/.

18. Pavlov, I. LZMA SDK (Software Development Kit). https://www.7-zip.org/sdk.html (2013).

19. Collet, Y. lz4. https://github.com/lz4.

20. Dean, J., Ghemawat, S. & Gunderson, S. H. snappy. https://github.com/google/snappy.

21. Masui, K. *et al.* A compression scheme for radio data in high performance computing. *Astronomy and Computing* **12**, 181–190 (2015).

22. Chee, S. W., Baraissov, Z., Loh, N. D., Matsudaira, P. T. & Mirsaidov, U. Desorption-Mediated Motion of Nanoparticles at the Liquid–Solid Interface. *J. Phys. Chem. C* **120**, 20462–20470 (2016).

23. Duane Loh, N. *et al.* Multistep nucleation of nanocrystals in aqueous solution. *Nat. Chem.* **9**, 77–82 (2016).

24. Koneti, S. *et al.* Fast electron tomography: Applications to beam sensitive samples and in situ TEM or operando environmental TEM studies. *Mater. Charact.* **151**, 480–495 (2019).

25. Jiang, Y. *et al.* Electron ptychography of 2D materials to deep sub-ångström resolution. *Nature* **559**, 343–349 (2018).

26. Ophus, C. Four-Dimensional Scanning Transmission Electron Microscopy (4D-STEM): From Scanning Nanodiffraction to Ptychography and Beyond. *Microsc. Microanal.* **25**, 563–582 (2019).

27. Pelz, P. M., Qiu, W. X., Bücker, R., Kassier, G. & Miller, R. J. D. Low-dose cryo electron ptychography via non-convex Bayesian optimization. *Sci. Rep.* **7**, 9883 (2017).

28. Hattne, J. *et al.* Analysis of Global and Site-Specific Radiation Damage in Cryo-EM. *Structure* **26**, 759–766.e4 (2018).

29. Clough, R. & Kirkland, A. I. Chapter One - Direct Digital Electron Detectors. in *Advances in Imaging and Electron Physics* (ed. Hawkes, P. W.) vol. 198 1–42 (Elsevier, 2016).

30. McLeod, R. A., Diogo Righetto, R., Stewart, A. & Stahlberg, H. MRCZ - A file format for cryo-TEM data with fast compression. *J. Struct. Biol.* **201**, 252–257 (2018).

31. Casañal, A. *et al.* Architecture of eukaryotic mRNA 3′-end processing machinery. *Science* vol. 358 1056–1059 (2017).

32. Falcon, B. *et al.* Novel tau filament fold in chronic traumatic encephalopathy encloses hydrophobic molecules. *Nature* **568**, 420–423 (2019).

33. Hofmann, S. *et al.* Conformation space of a heterodimeric ABC exporter under turnover conditions. *Nature* **571**, 580–583 (2019).

34. Zhao, P. *et al.* Activation of the GLP-1 receptor by a non-peptidic agonist. *Nature* **577**, 432–436 (2020).

35. Vincent, L. & Soille, P. Watersheds in digital spaces: an efficient algorithm based on immersion simulations. *IEEE Trans. Pattern Anal. Mach. Intell.* 583–598 (1991).

36. McMullan, G., Faruqi, A. R. & Henderson, R. Direct Electron Detectors. *Methods Enzymol.* **579**, 1–17 (2016).

37. Ruskin, R. S., Yu, Z. & Grigorieff, N. Quantitative characterization of electron detectors for transmission electron microscopy. *J. Struct. Biol.* **184**, 385–393 (2013).

# Figure Captions

Figure 1: **(a)** Data reduction levels and scheme. The leftmost image (L0) depicts a 10x10 pixel image (the raw detector output) with four secondary electron puddles. The remaining four images from left to right correspond to the four data reduction levels, L1 to L4, respectively. Each image represents a reconstruction of the original image (L0) using only the information retained at that level (see table at the bottom). The L1 image retains all the useful information about the secondary puddles by first removing detector readout/thermal noise from L0. In L2, the spatial location of the four puddles, the number of pixels (area) in each puddle, the shape of the four puddles and an intensity summary statistic (sum, maximum or mean) for each puddle are retained. Each reduction level offers different advantages in terms of speed, compression, information loss, spatial or temporal resolution, etc (see row labelled "Optimized For"). The row labeled "Reduced Representation" describes how the information retained at each level is packed in the reduced format. These packings are tuned to provide a good balance between reduction speed and compressibility. In L3, the puddle area, shape and location information are all encoded in a single binary image, which is easily computed and highly compressible. These three aspects in L1 and L2 are packed as the binary image used in L3. Only the most likely locations of incident electrons are saved as binary maps in L4. Panels **(b)**, **(c)**, **(d)** and **(e)** are the reduction compression pipelines for reduction levels L1, L2, L3 and L4, respectively. Here, the thresholding step produces a binary map identifying pixels as signal or noise. Bit packing removes unused bits and converts the list of ADU values into a continuous string of bits. The connected components labelling algorithm identifies clusters of connected pixels that constitute individual electron puddles from this binary map. Puddle centroid extraction further reduces each puddle to a single representative pixel; and puddle feature extraction computes puddle specific features such as mean or maximum ADU.

Figure 2: Recalibration of L1 reduced data to remove artifacts. Panels (a) and (b) are Fourier transforms (FT) of summed L1 reduced frames of HRTEM movies of a molybdenum disulfide 2-D crystal, acquired using a JEOL 2200 microscope operating at 200keV and a DE-16 detector running at 300 fps, with a pixel resolution of 0.2 Å (a) is L1 reduced using fast on-the-fly calibration using a $3\sigma$ threshold (see Methods Section) (b) is the result of recalibrating (a) with a more stringent fine calibration that uses an area threshold and a $4\sigma$ threshold (see Methods Section). The Fourier peaks indicated with orange arrows in (a) are due to detector artifacts, which are not readily visible in the image but can severely impact drift correction. (a) and (b) are the sum of FFTs of 9000 frames.

Figure 3: Reducibility and compressibility of data with increasing electron flux. The solid black line ("unreduced compression") shows the compression ratios achieved on unreduced raw data (including dark noise) using Deflate-1. The dashed lines show the compression ratios achieved with just the four levels of data reduction and without any compression. The solid lines show the compression ratios after compressing the reduced data using Deflate-1. The coincidence loss levels corresponding to the electron fluxes label the second $y$-axis on the right.

Figure 4: Comparison of compression algorithms with L1 reduction at three dose rates. Each scatter plot shows the reduction compression ratios and the compression throughputs of six compression algorithms (Deflate, Zstandard (Zstd), bzip2 (Bzip), LZ4, LZMA, and SNAPPY), plus the Blosc variants of Deflate, Zstandard (Zstd),

LZ4, and SNAPPY. Reduction compression ratio (horizontal axes in all panels) is the ratio between the raw (uncompressed) data and the reduced compressed data sizes. The three rows of scatter plots correspond to three different electron fluxes: 0.01, 0.03 and 0.05 e/pixel/frame, from top to bottom. The left and right columns of scatter plots correspond to the two most extreme internal optimization levels of the compression algorithms: fastest but suboptimal compression labeled "Optimal Speed" (left column), and optimal but slow compression labeled "Optimal Compression" (right column). The data throughputs (vertical axes in all panels) are based on single threaded operation of ReCoDe and include the time taken for both reduction and compression. The decompression throughputs of the six algorithms are presented in Supplementary Fig. S2.

Figure 5: Pipeline and data throughput of on-the-fly reduction and compression. **(a)** ReCoDe's multithreaded reduction compression pipeline used for live data acquisition. The CMOS detector writes data into the RAM-disk in timed chunks, which the ReCoDe server processes onto local buffers and then moves to NAS servers. The ReCoDe Queue Manager synchronizes interactions between the ReCoDe server and the detector. **(b)** L1 reduction and compression throughput (GB/s) of Deflate-1, with multiple cores at four electron fluxes. The throughput of ReCoDe depends only on the number of electron events every second, hence the four dose rates (horizontal axis) are labelled in million electrons/second. The simulations were performed on a 28-core system, as a result, throughput scales non-linearly when using more than 28 cores (Supplementary Figure S3). **(c)** and **(d)** show throughputs when using 10 GbE and IPoIB connections to write directly to NAS, respectively. In **(e)**, throughput of L1 reduction without any compression; **(f)** throughput of Deflate-1 when compressing the unreduced raw data. **(g)** shows the conversion between million e/s and e/pixel/frame for two different frame size-frame rate configurations of the DE-16 detector.

Figure 6: Maximum data acquisition time of 1TB of movie-mode TEM without data reduction and compression. Each cell's horizontal and vertical grid position marks the temporal resolution (or, equivalently, frame rate) and frame size of a hypothetical movie-mode data acquisition scenario respectively. A cell's text and color indicates the time taken to acquire one terabyte (TB) of data at that frame size and temporal resolution without reduction and compression. For larger frames and high temporal resolution (top right corner), acquisitions lasting merely tens of seconds already produce 1 TB of data. With a 95x reduction in data size the same experiment can span 20 times longer, enabling the observation of millisecond dynamics in reactions that span several minutes. The yellow dots show a few of the frame size-frame rate combinations available for the DE-16 detector.

Figure 7: Comparison of ReCoDe and MRCZ for archival datasets in EMPIAR. (a) shows that the compression ratios obtained by ReCoDe (filled stars) on relatively low dose rate EMPIAR datasets are higher than those due to MRCZ (filled circles). (b) Compression ratios obtained using MRCZ and ReCoDe on simulated 16-bit unsigned integer data. The crossover point for performance occurs at 0.58 electron/pix/frame. At dose rates below this ReCoDe achieves higher compression ratios than MRCZ, whereas at dose rates above this MRCZ achieves slightly higher compression ratios. The electron events per pixel follows a Poisson distribution in these simulated datasets. The underlying compression algorithms used in (a) and (b) is Blosc + Deflate (zlib) for both MRCZ and ReCoDe. (c) lists a short description of the seven EMPIAR dataset used to generate (a). Overall in the simulated data, for both compression algorithms, compression ratios reduce as dose rate increases, as expected. However, for the EMPIAR datasets, there are two groups, one for the floating-point data (datasets 0-5) and another for integer data (datasets 6 and 7). Although the floating-point data have lower dose rates than the integer type data, the former is less compressible because they are naturally less sparse than the latter. Nevertheless, within each group, the expected trend (reduction in compression ratio with increasing dose rate) holds true and ReCoDe outperforms MRCZ. A comparison where all the datasets are standardized to the same integer data type, presented in Supplementary Figure S6, shows that the results from EMPIAR datasets and simulated data are quite similar.

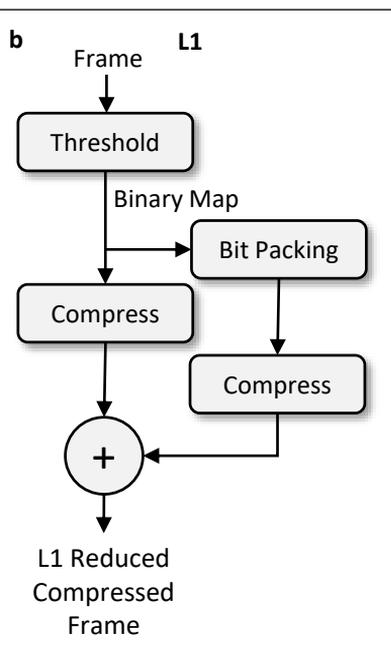
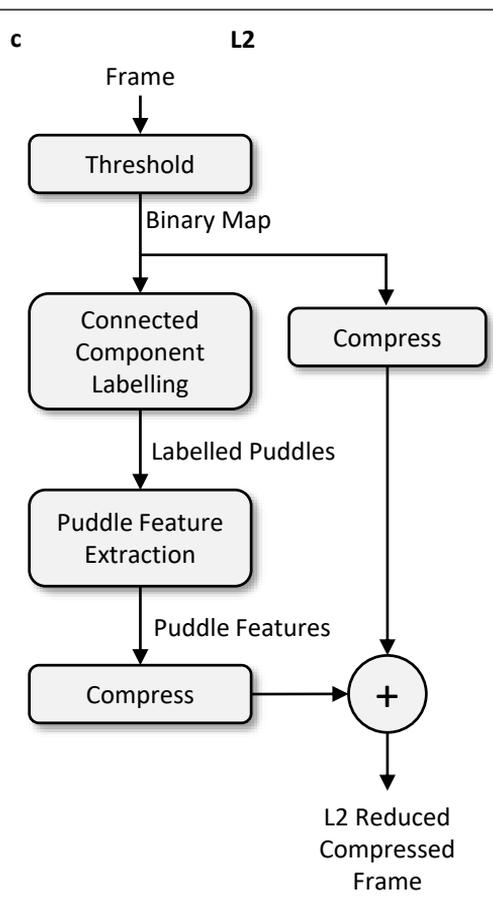
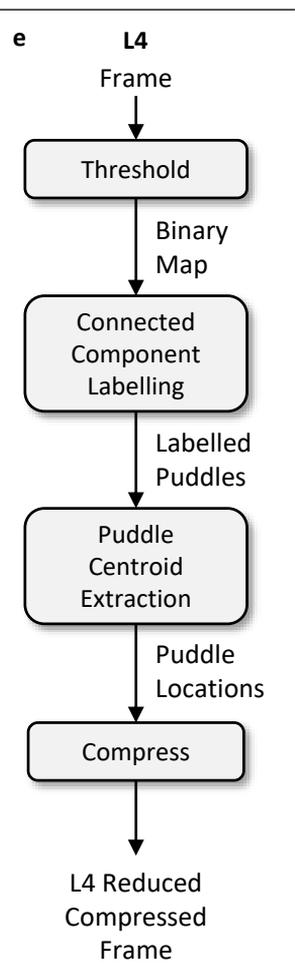
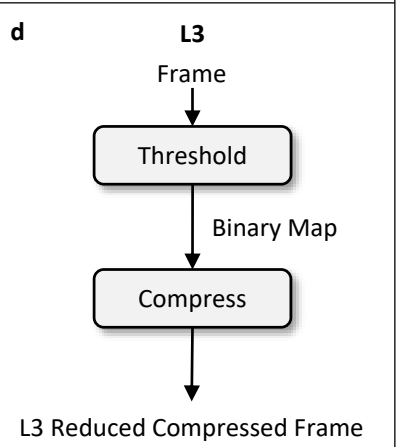

Figure 1

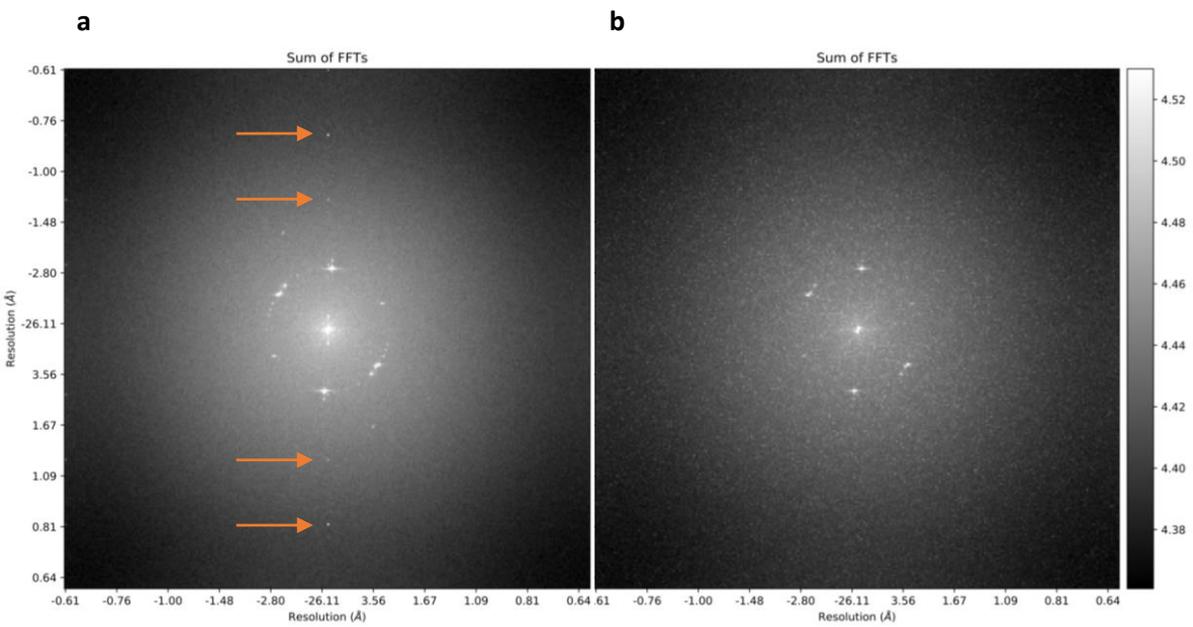

Figure 2

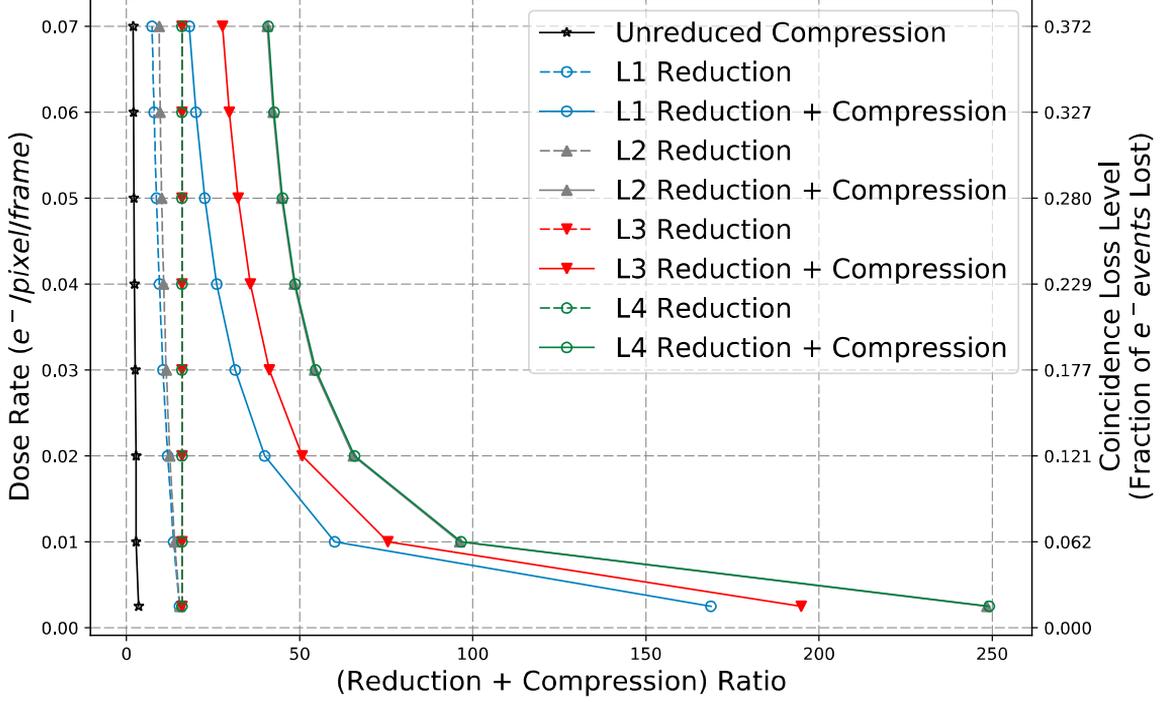

Figure 3

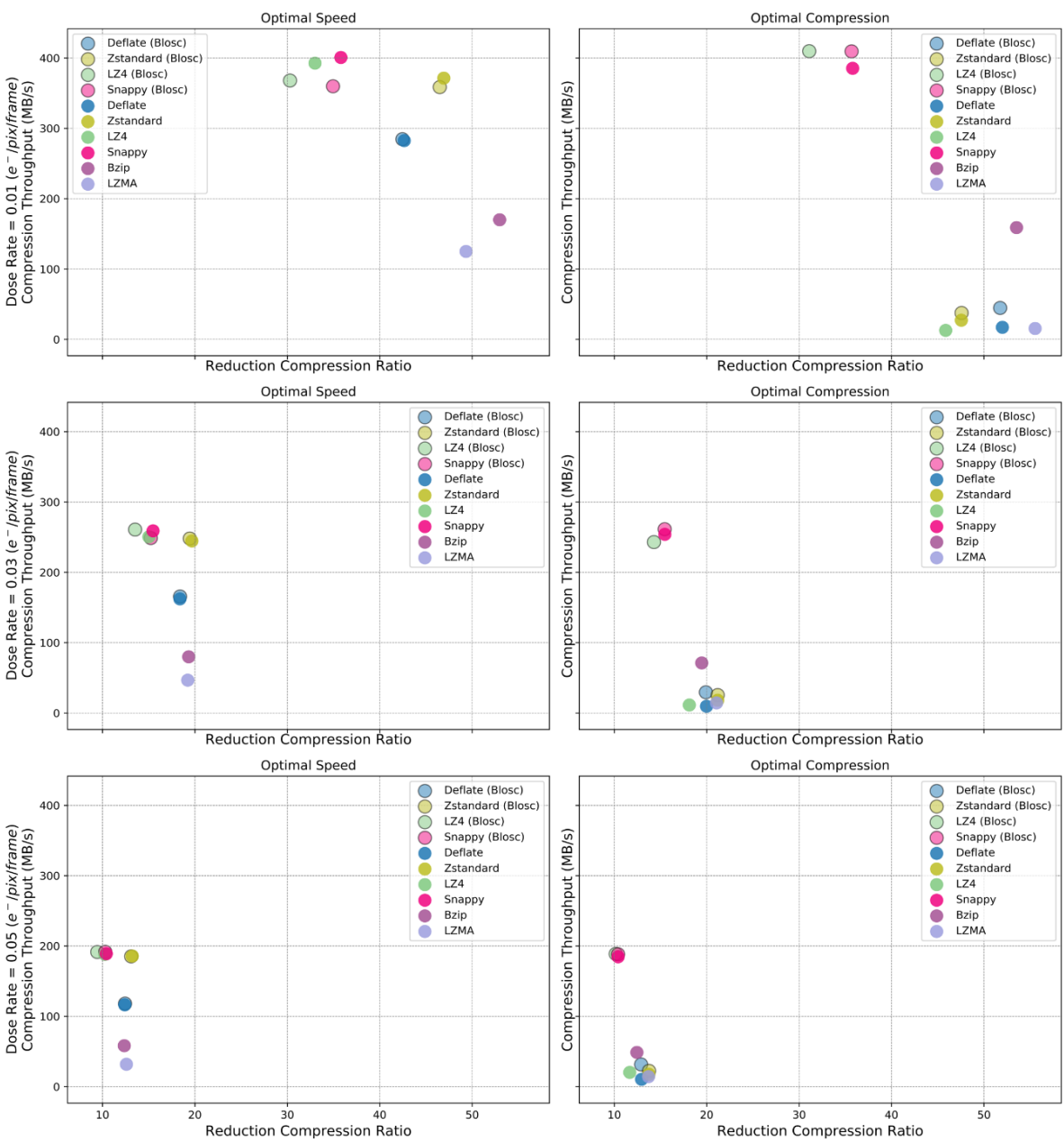

Figure 4

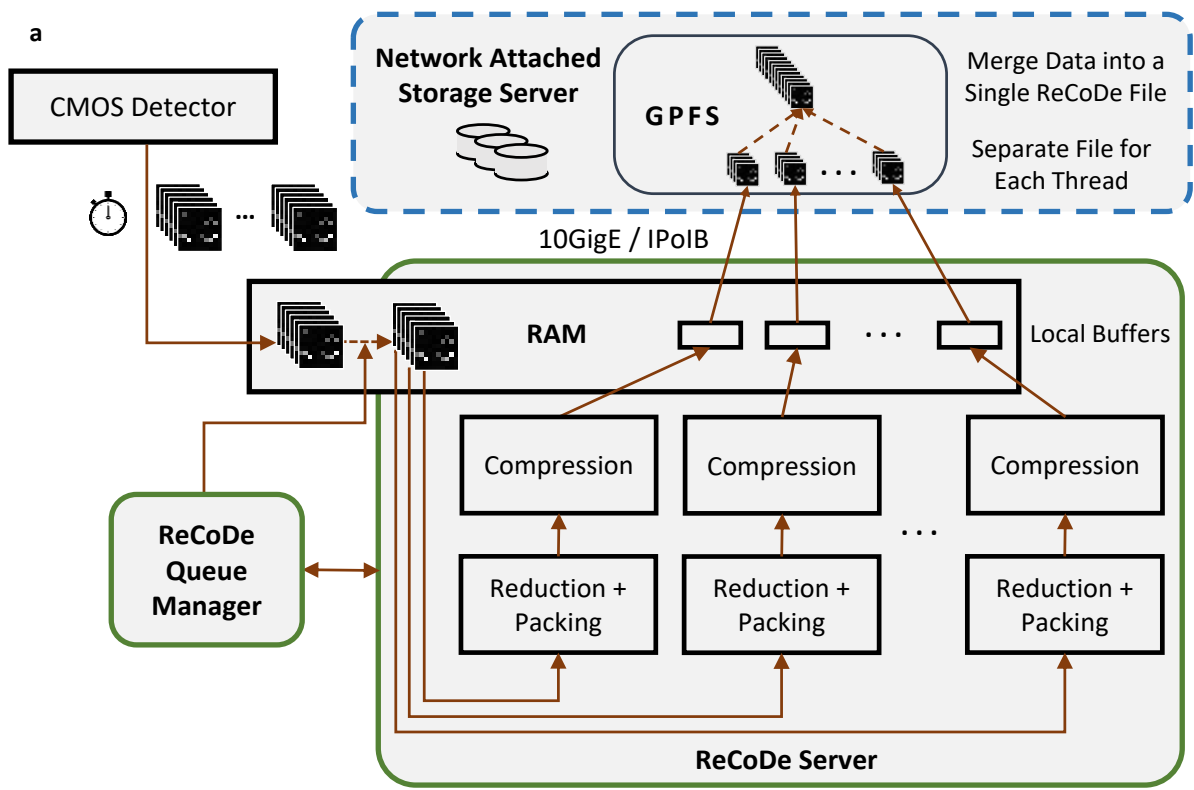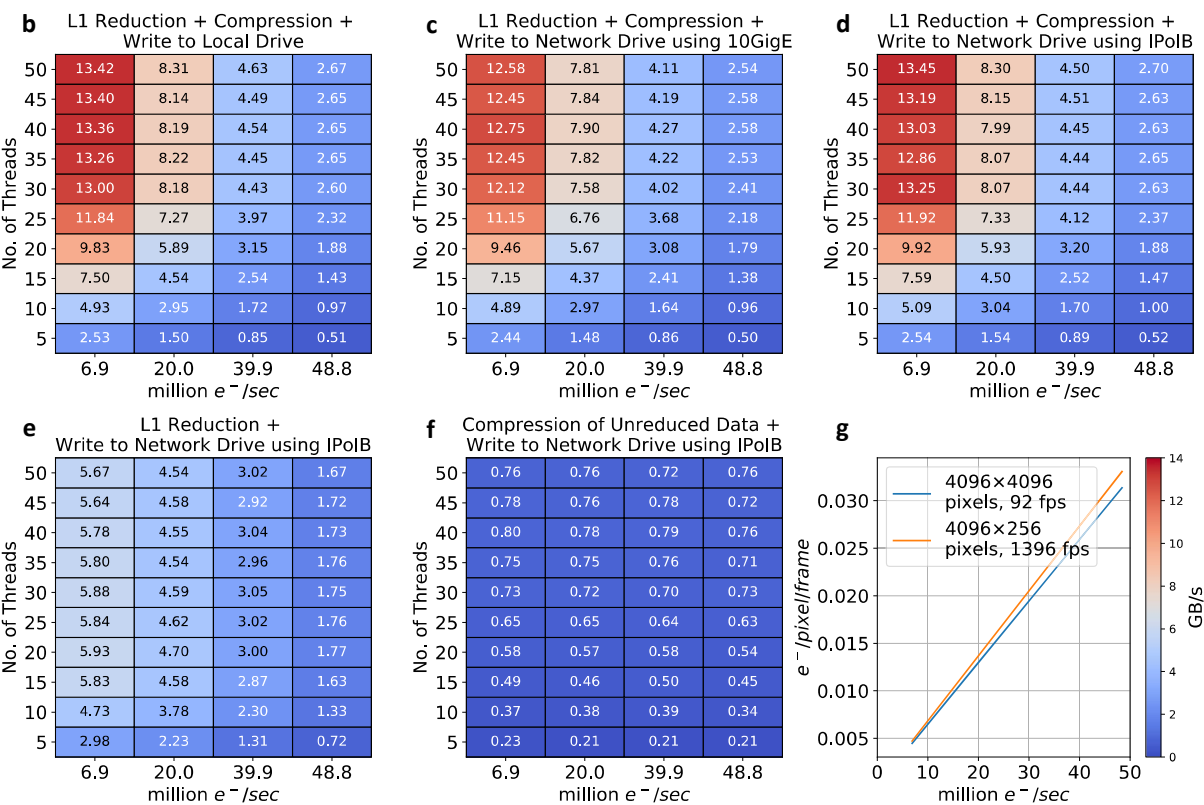

Figure 5

# Table 1

| Dose Rate ($e^-$/pixel/frame) | Coincidence Loss (Fraction of $e^-$ events Lost) | | | | |
|---|---|---|---|---|---|
| | Simulated with fixed Size and Shape | | | Analytically Computed using Size Distribution | Simulated using Size and Shape Distributions |
| | Assuming 3 × 3 pixel PSF | Assuming 2 × 2 pixel PSF | Assuming 1 pixel PSF | | |
| 0.0025 | 0.060 | 0.031 | 0.011 | 0.0140 | 0.0156 |
| 0.005 | 0.117 | 0.061 | 0.022 | 0.0354 | 0.0317 |
| 0.01 | 0.725 | 0.119 | 0.044 | 0.0784 | 0.0619 |
| 0.02 | 0.407 | 0.227 | 0.087 | 0.1524 | 0.1207 |
| 0.03 | 0.556 | 0.324 | 0.128 | 0.2115 | 0.1767 |
| 0.04 | 0.676 | 0.413 | 0.167 | 0.2590 | 0.2296 |
| 0.05 | 0.772 | 0.492 | 0.205 | 0.2975 | 0.2801 |
| 0.06 | 0.846 | 0.563 | 0.242 | 0.3288 | 0.3272 |
| 0.07 | 0.902 | 0.626 | 0.278 | 0.3544 | 0.3726 |
| 0.08 | 0.942 | 0.684 | 0.312 | 0.3753 | 0.4155 |
| 0.09 | 0.969 | 0.734 | 0.345 | 0.3923 | 0.4558 |
| 0.1 | 0.983 | 0.779 | 0.376 | 0.4061 | 0.4940 |

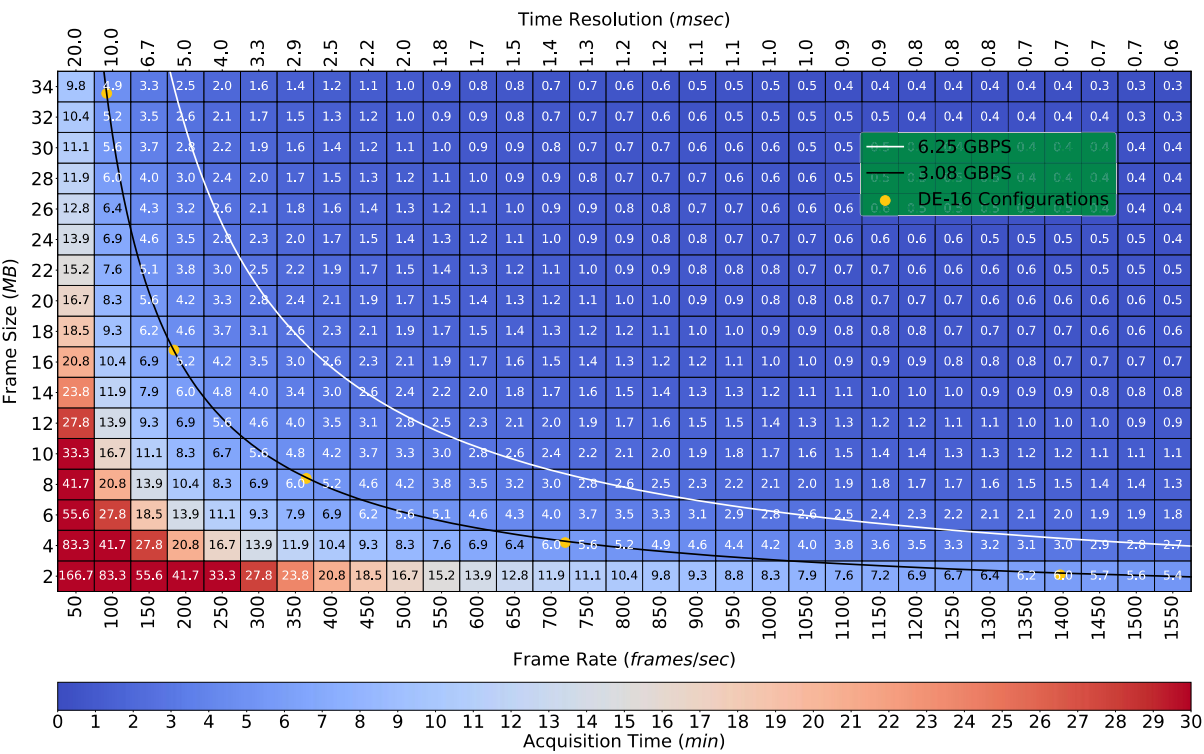

Figure 6

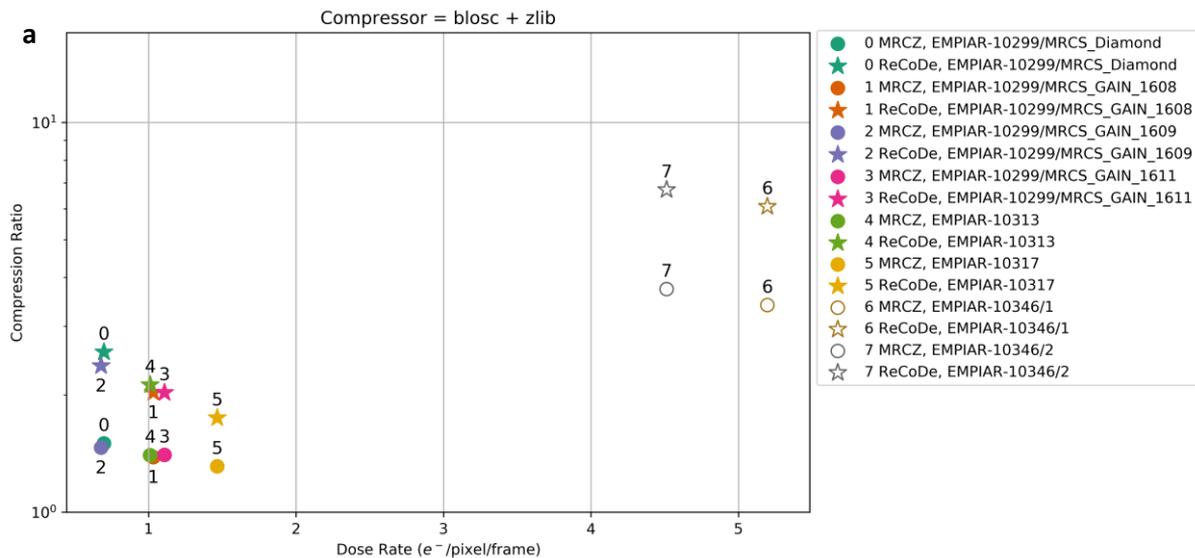

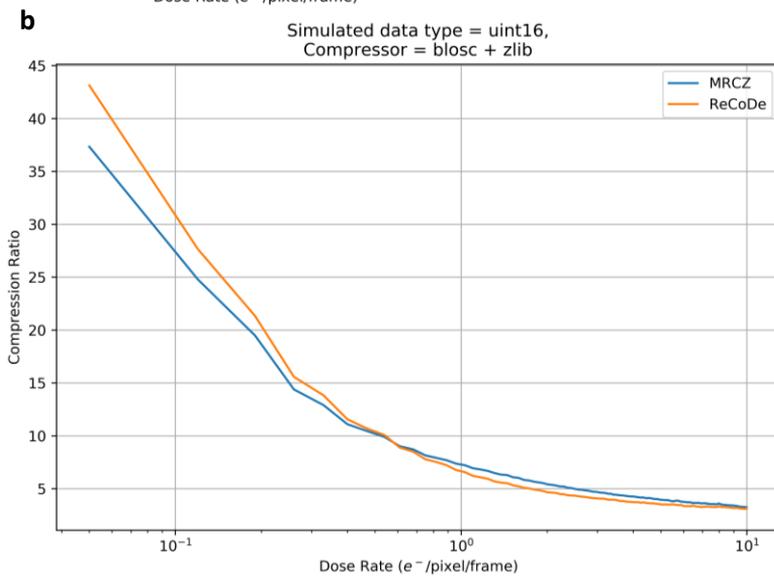

| No. | Dataset | Dose Rate (e/p/s) | Detector | Image Size | Data Type | Reference |
|---|---|---|---|---|---|---|
| 0 | 10299/MRCS_Diamond | 0.69 | K2 Summit | 7676 x 7420 | Float 32-bit | Casañal et al. |
| 1 | 10299/MRCS_GAIN_1608 | 1.03 | Falcon II | 7676 x 7420 | Float 32-bit | |
| 2 | 10299/MRCS_GAIN_1609 | 0.67 | Falcon II | 7676 x 7420 | Float 32-bit | |
| 3 | 10299/MRCS_GAIN_1611 | 1.107 | Falcon II | 7676 x 7420 | Float 32-bit | |
| 4 | 10313 | 1.01 | K2 Summit | 3838 x 3710 | Float 32-bit | Falcon et al. |
| 5 | 10317 | 1.46 | K2 Summit | 3838 x 3710 | Float 32-bit | Hofmann et al. |
| 6 | 10346 | 5.19 | K3 | 11520 x 8184 | Unsigned Byte | Zhao et al. |
| 7 | 10346 | 4.51 | K3 | 11520 x 8184 | Unsigned Byte | |

Figure 7

# Supplementary Information on "ReCoDe: A Data Reduction and Compression Description for High Throughput Time-Resolved Electron Microscopy"

Abhik Datta[1,2], Kian Fong Ng[1,2], Balakrishnan Deepan[1,2], Melissa Ding[3], See Wee Chee[1,2,4], Yvonne Ban[1,2], Jian Shi[1,2], Duane Loh[1,2,4]

[1]Centre for BioImaging Sciences, National University of Singapore, Singapore 117557.
[2]Department of Biological Sciences, National University of Singapore, Singapore 117557.
[3]Department of Computer Science and Engineering, Ohio State University, Columbus, OH 43210, USA.
[4]Department of Physics, National University of Singapore, Singapore 117551.

## Table of Contents

# ReCoDe Data Format

## Supplementary Method S1: The ReCoDe Data Format

Reduced Compressed Data Format

All intermediate and merged ReCoDe files begin with a ReCoDe header which has a fixed length (512 bytes) and a static structure (described in Supplementary Table S1). The headers are followed by an optional section (non-standard frame metadata descriptions) that lists names and sizes of additional fields, such as timepoint or scan position, in each frame's metadata. This is followed by the header of the original file (such as MRCS or Sequence headers). In merged files, these are followed by a frame metadata section that stores the compressed sizes of frame data as well as the values of additional non-standard frame metadata. This section is followed by the actual reduced compressed frame data. The merged ReCoDe file structure is described in Supplementary Fig. S1a. Optionally, the per-pixel calibration data, used for signal-noise separation, can be appended at the end of the file as a single frame. The availability of this frame is indicated in the ReCoDe header field "Is Calibration Data Appended" (see Supplementary Table S1).

The intermediate files generated during multithreaded reduction compression are optimized for sequential access and have a slightly different structure than merged ReCoDe files. In intermediate files, the ReCoDe header, the non-standard frame metadata descriptions, and the original file's header are followed by frame-blocks, with each frame-block containing a single frame's metadata followed by the actual reduced compressed data for the frame. The intermediate ReCoDe file structure is described in Supplementary Fig. S1b.

The exact frame metadata information and the reduced compressed information retained for each frame depends on the reduction level. For example, in L1 the metadata for each frame contains three 4-byte-long numbers: the size of the compressed binary image, the size of the compressed pixel intensities, and the number of foreground pixels. Retaining the size of compressed data in each frame is necessary as the compressed sizes vary. Storing these sizes in the metadata block, which appears before the actual compressed frame data in the merged file, allows efficient random access of frames in merged files by only parsing the metadata block. In contrast, the intermediate files are designed for sequential data access. Decompressing the pixel intensity values requires knowing the size of the decompressed data, therefore storing the number of foreground pixels explicitly rather than inferring it from the decompressed binary map, allows the binary map and the pixel intensity values to be decompressed in parallel. The frame metadata section in intermediate files additionally stores frame ids. Since frames may not be distributed to intermediate files in any specific order, frame ids are used to ensure that frames are placed sequentially when creating the merged file in the absence of other optional frame ordering information such as timepoints.

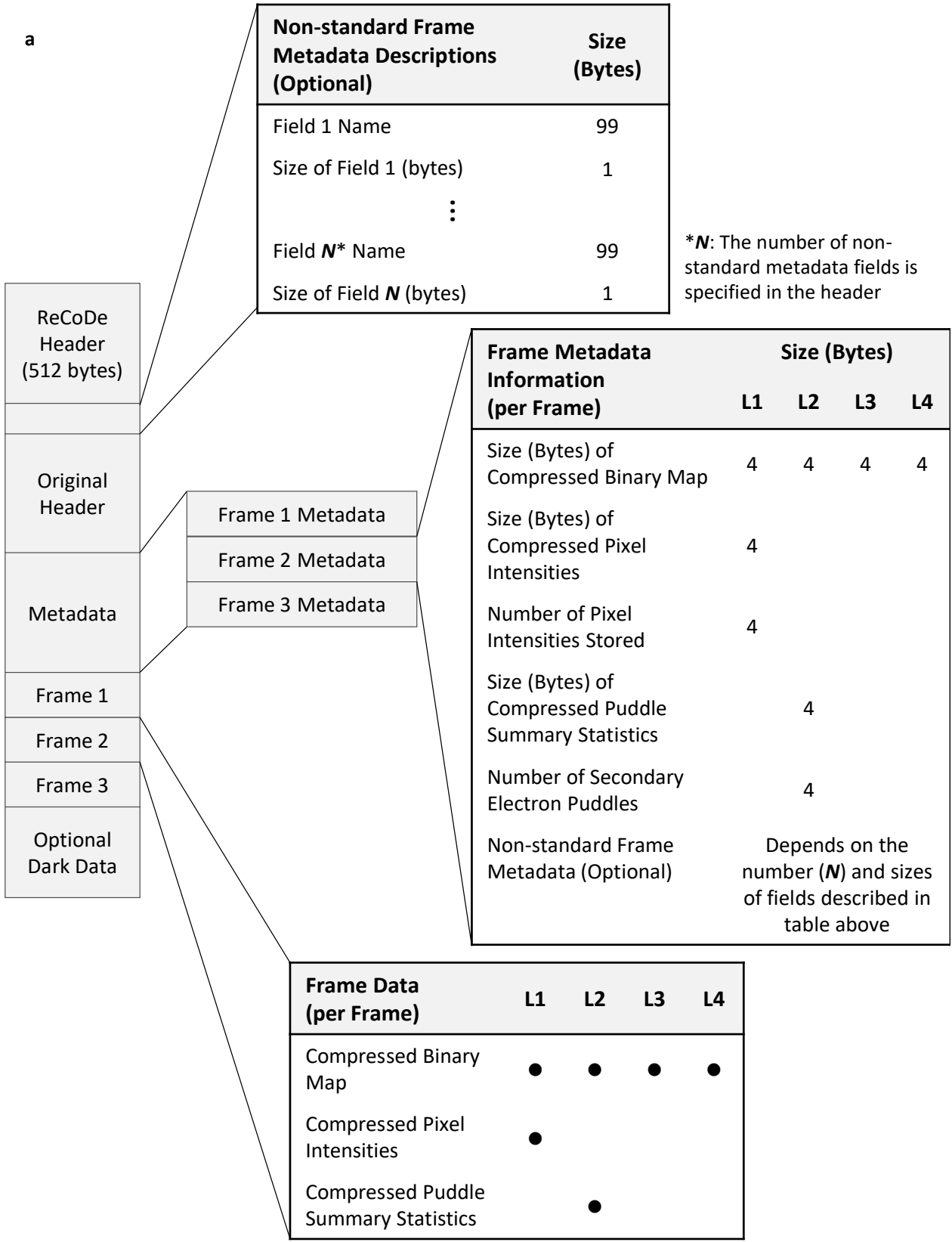

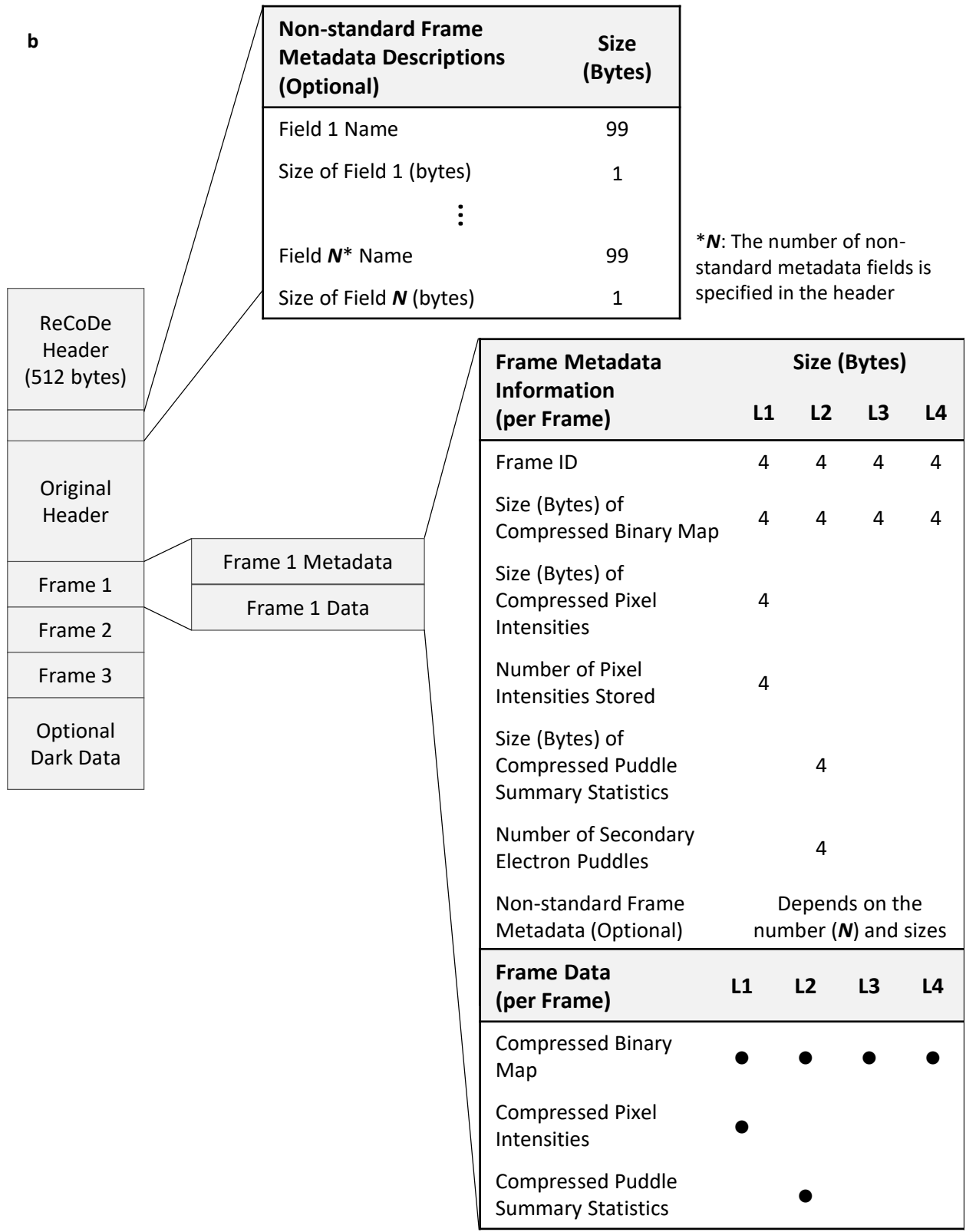

Supplementary Figure S1: ReCoDe file structure for a time series with only 3 frames. (**a**) Structure of merged ReCoDe file and (**b**) structure of intermediate ReCoDe files.

Supplementary Table S1: Structure of ReCoDe header.

| Field | Size (bytes) | Possible Values (Code = Value) / Description |
|---|---|---|
| Unique Identifier | 8 | Always 158966344846346 |
| Major Version | 1 | |
| Minor Version | 1 | |
| Is Intermediate File | 1 | 1 = Intermediate File, 0 = Merged File |
| Reduction Level | 1 | 1 - 4 |
| Is Compressed | 1 | 0 = Reduction Only, 1 = Reduction + Compression |
| Is Bit-packed | 1 | 0 = False., 1 = True |
| Target Intensity Bit-depth | 1 | Bit-depth of output (ReCoDe) data |
| Number of Rows In a Frame | 4 | |
| Number of Columns in a Frame | 4 | |
| Number of Frames | 4 | |
| Frame Metadata Size | 1 | Size of each frame's metadata in bytes |
| Number of Non-standard Frame Metadata Fields | 1 | No. of additional frame specific information (time point, scan position, etc.) |
| Secondary Electron Puddle Statistics | 1 | 0 = Default (Max), 1 = Max, 2 = Sum. Used only in L2. |
| Secondary Electron Puddle Centroiding Scheme | 1 | 0 = Default (Centre of Mass), 1 = Centre of Mass, 2 = Max. Pixel, 3 = Central Pixel. Used only in L2 and L4. |
| Compression Method | 1 | 0 = Deflate, 1 = bzip, 2 = LZMA, 3 = Snappy, 4 = Lz4 |
| Compression Level | 1 | 1 – 9. Internal optimization level of compression algorithms |
| Original File Type | 1 | 0 = Binary, 1 = MRC, 2 = Sequence, 255 = Others |
| Original File Header Size | 2 | 1024 for MRCS and Sequence, 0 otherwise |
| Original File Header Position | 1 | Always 1 (indicating after ReCoDe header). The option "before ReCoDe header" is deprecated in version 0.2. |
| Original Source File Name | 100 | |
| Dark Noise File Name | 100 | |
| Signal-Noise Calibration Parameter (S) | 8 | See signal-noise calibration algorithm for details. Assumed to have the same data type as source data. |
| Is Calibration Data Appended | 1 | 0 = False, 1 = True |
| Frame Offset | 4 | Index of First Used Frame in Original Data |
| Dark Frame Offset | 4 | Index of First Used Frame in Calibration Data |
| No. of Dark Frames Used for Calibration | 4 | |
| Source Bit-depth | 1 | Bit-depth of original data. Can be different from output bit-depth, if scaling is used. |
| Source Data Type | 1 | 0 = Unsigned Integer, 1 = Signed Integer, 2 = Float |
| Target Data Type | 1 | Data type of output (ReCoDe) data. 0 = Unsigned Integer, 1 = Signed Integer, 2 = Float |
| Checksum | 32 | |
| Reserved For Future Use | 219 | |

Reduced Data Format

ReCoDe also has the provision of only reducing the data without compressing it. The reduced data files (merged and intermediate) follow the same structure as the corresponding reduced and compressed ReCoDe files. However, the frame metadata sections for L1 and L2 reduction levels, stores the number of pixel intensities and the number of secondary electron puddles, respectively, in addition to any non-standard frame metadata. For L3 and L4 reduction levels, there is no standard metadata information, only non-standard frame metadata are stored if present. The frame data section stores the same information as in the reduction compression case but in uncompressed form.

Validation frames

ReCoDe also has the option of saving every $n^{th}$ frame (where $n$ is user-specified) simultaneously in an uncompressed unreduced raw format. These frames are saved as a separate MRC stack and a reduced-compressed version of this frame is still saved in the ReCoDe format. Comparing these two versions of the same frame can serve as a useful validation/diagnostic.

Merging Part Files

Frames processed by a reduction compression thread are sequentially appended to its intermediate file. If the frames processed by a reduction compression thread are time-ordered, i.e. the $(i-1)^{th}$ frame precedes the $i^{th}$ frame in time, the merging step becomes trivial. Merging $k$ intermediate files require maintaining a pointer to the earliest unmerged frame in each intermediate file. The merging involves finding the earliest frame among the frames pointed to by the $k$ pointers (say the $i^{th}$ frame of the $n^{th}$ intermediate file), adding that frame to the merged file, and moving the $n^{th}$ pointer to the next frame in the intermediate file. This merging accesses the intermediate files and the merged output file sequentially and is therefore very fast.

# Data Compression

## Supplementary Figure S2: Decompression Throughput

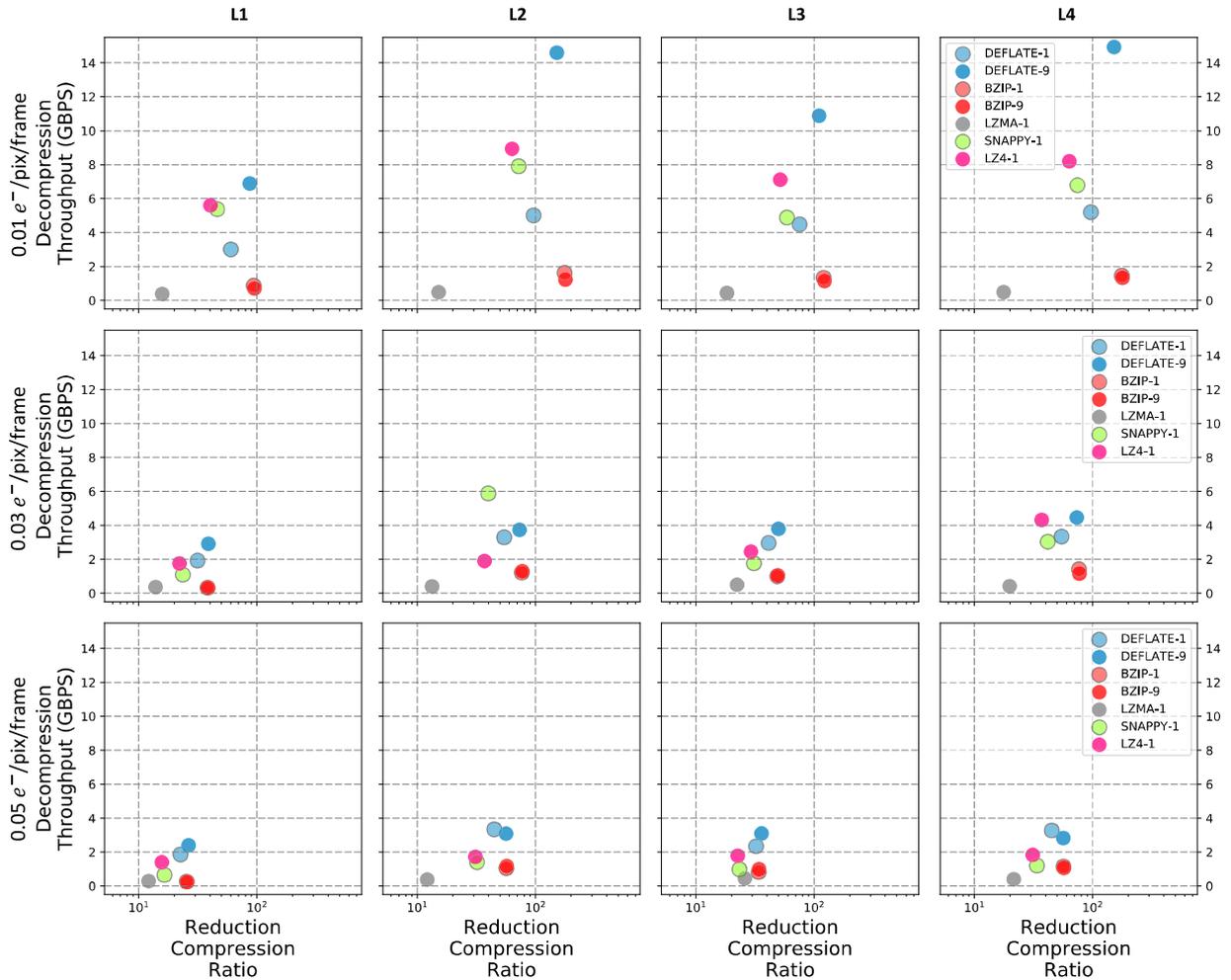

Supplementary Figure S2: Decompression throughputs. The reduction compression ratios and decompression throughputs of five algorithms: Deflate, bzip2, LZ4, LZMA and SNAPPY. Here, the reduction compression ratio is the ratio between the sizes of the original (uncompressed) data and the reduced compressed data. Suffix -1 and -9 refer to internal optimization levels of the algorithms corresponding to the fastest compression and the optimal compression, respectively. For LZ4, LZMA and SNAPPY the internal optimizations produced similar results, hence only the default level is shown. The three rows of scatter plots correspond to three different electron fluxes: 0.01, 0.03 and 0.05 e/pixel/frame from top to bottom, and the four columns of scatter plots correspond to the four reduction levels: L1 to L4 from left to right. The throughputs are based on single threaded operation of ReCoDe and include the time taken for both reduction and decompression.

## Supplementary Discussion S3: Single-Threaded Write Performance

In the absence of distributed storage servers, on-the-fly reduction compression will have to write to standard hard drives or SSDs, which support a limited number of simultaneous writes. To evaluate ReCoDe's on-the-fly reduction compression performance in such cases, we evaluated an alternative implementation, where the reduction compression threads do not independently write to their respective intermediate files. Instead, a single offloader thread performs all writes to disk. This offloader thread is invoked by the reduction-compression threads when their local buffers are full. The reduction compression threads then have to wait for the offloader thread to attend to their requests and clear the buffer. This implementation emulates the worst-case write performance, where a single thread sequentially accesses the disk. The experiment was repeated for four flux levels (Supplementary Fig. S3). From the observed speed-ups, we estimate, based on Amdahl's law[1], that even with one offloader thread ~89% of the code runs in parallel.

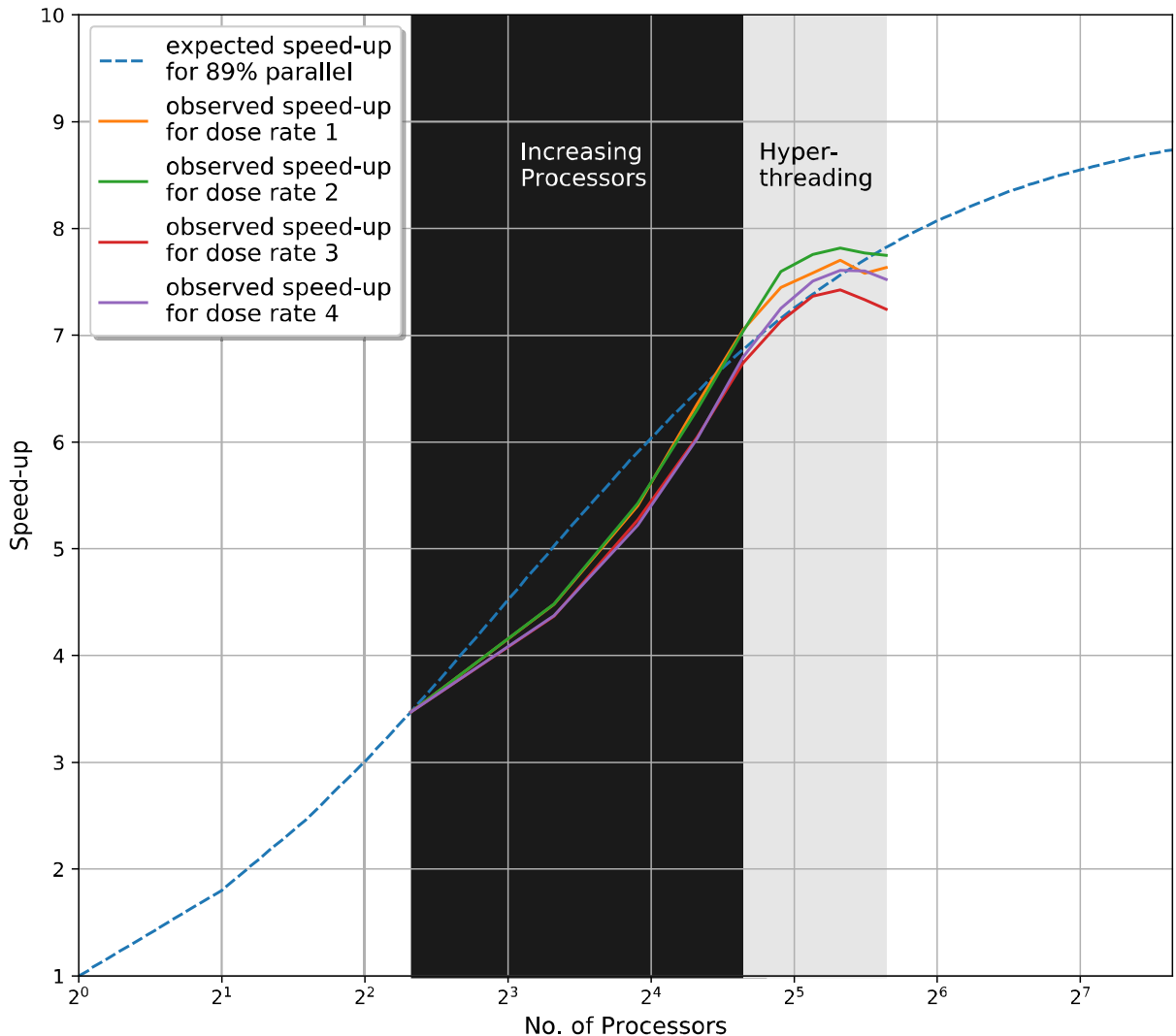

Supplementary Figure S3: Single threaded write performance of ReCoDe. The solid lines show the observed speed-ups with increasing processors for four flux levels: 0.005, 0.014, 0.027 and 0.034 e/pixel/frame. The blue dashed line shows the predicted speed-up for an 89% parallel code, as per Amdahl's law[1]. The simulations were performed on a single compute node with 28-cores (14 core x 2 chips), 2.6 GHz Intel Xeon processors and 512 GB RAM. Simulations using greater than 28 threads used hyperthreading.

## Supplementary Discussion S4: Effect of Data Reduction on Image Quality

A knife-edge test was performed using a beam blanker, with an electron flux of ~0.8 e/pixel/s using the DE-16 detector operating at 400 fps exposed for 50 seconds, resulting in a total dose of ~40 e/pixel and an effective electron flux of *~0.002*e/pixel/frame across 20,000 frames. Each frame was resampled to twice the resolution using bicubic interpolation and counting was performed on the resampled frames following one of the three localization strategies: for each electron puddle, we either localized to the weighted centroid, non-weighted centroid, or the pixel with the maximum value. The straight edge of the beam blanker was visually determined in the counted image obtained by summing the counted frames. The edge spread function (ESF) was estimated from the average of multiple one-dimensional profiles measured along the normal to the straight edge. The edge spread function was differentiated to give the line spread function (LSF) which was Fourier transformed to get the MTF.

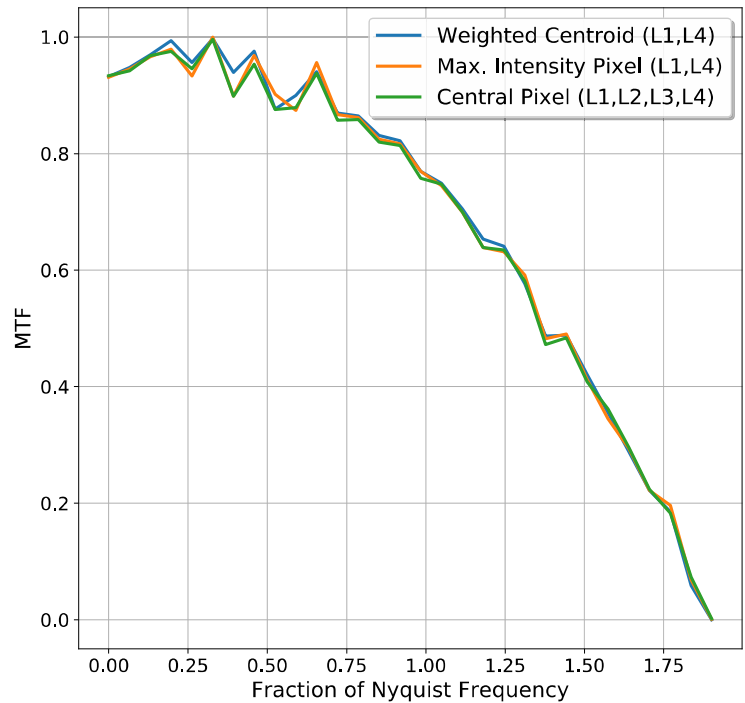

Supplementary Figure S4: Data reduction and image quality. MTFs estimated using the knife-edge method from three counted images of the beam blanker are shown. The three counted images are obtained using three different approaches to estimating the entry point of electrons from their secondary electron puddles (see Methods section for implementation details). For this MTF calculation a Dirac delta function is used as the PSF (instead of the actual detector PSF) to highlight the differences due to localization errors alone.

## Supplementary Discussion S5: Coincidence Loss Estimation

To understand the role of puddle shape and size in coincidence loss estimation, five coincidence loss estimation strategies were evaluated (Table 1 of Main text). The first strategy assumes all secondary electron puddles to be of the same shape and size: *3 × 3* pixels. With these assumptions, simulated images were used to calculate the actual frequency of puddle overlaps at twelve different dose rates. Similarly, in the second and third strategies puddle of sizes *2 × 2* pixels and 1 pixel were used to simulate images, respectively. In the fourth strategy, the size distribution of puddles was learned from actual DE-16 detector data, collected at very low dose rates (0.001 e/pixel/frame). The puddles were, however, assumed to be circular and the expected coincidence loss was analytically computed. The final strategy was to simulate images following the learned shape as well as size distributions of the puddle and compute the actual frequency of puddle overlaps. The results indicate that shape and size information is critical in estimating coincidence loss, without which, coincidence loss values are underestimated (third and fourth columns compared to the fifth column in Table 1 of Main text).

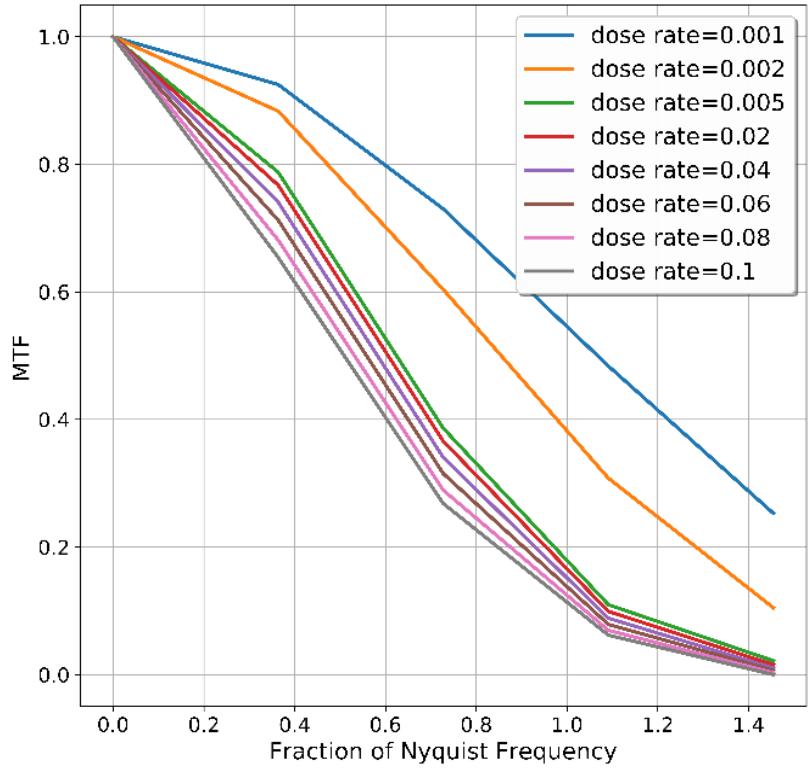

Supplementary Figure S5: Effects of coincidence loss on counting accuracy. MTFs corresponding to counted images, simulate at electron fluxes ranging from 0.005 to 0.1 $e/Å^2/s$ are shown. As electron flux increases, the MTF of counted images decrease due to higher coincidence loss, particularly at higher frequencies. The counting used here reduces each puddle to its weighted centroid pixel.

# Supplementary Discussion S6: Storage Requirements of Cryo-EM Experiments with Movie-mode Detectors

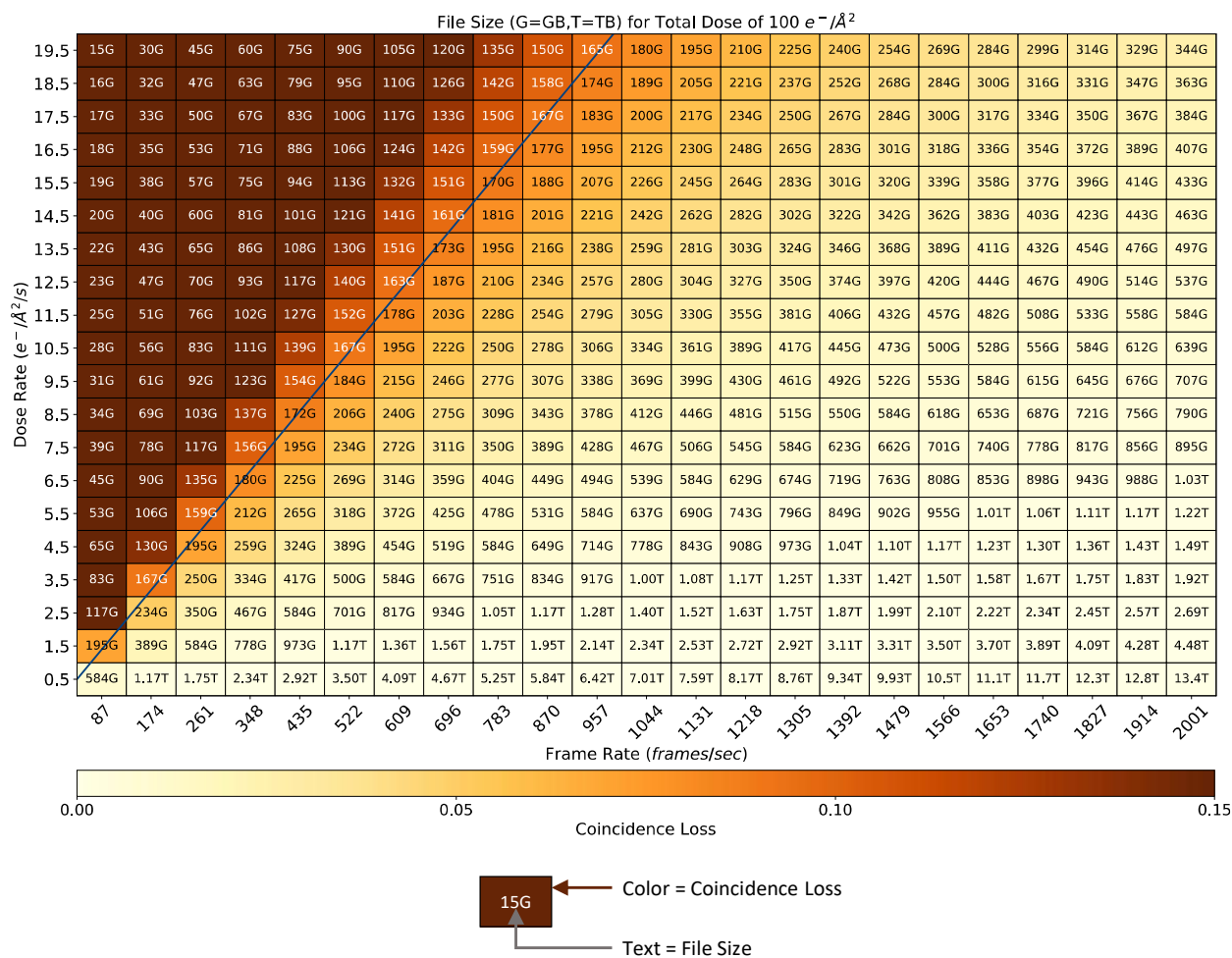

Supplementary Figure S6: Storage requirements of cryo-EM experiments with movie-mode detectors. Shows how lower dose rates and higher temporal resolutions require collecting prohibitively large amounts of data in the absence of reduction compression. The text in each cell indicates the amount of data needed to achieve a total dose of 100 e/$Å^2$/s, across different electron fluxes and temporal resolutions. This calculation assumes a 4096×4096 pixel detector producing 16-bit images at a magnification with pixel size equal to $1Å$. The cell colors represent the coincidence loss suffered at that electron flux and temporal resolution.

# Supplementary Figure S7: Comparison of ReCoDe and MRCZ

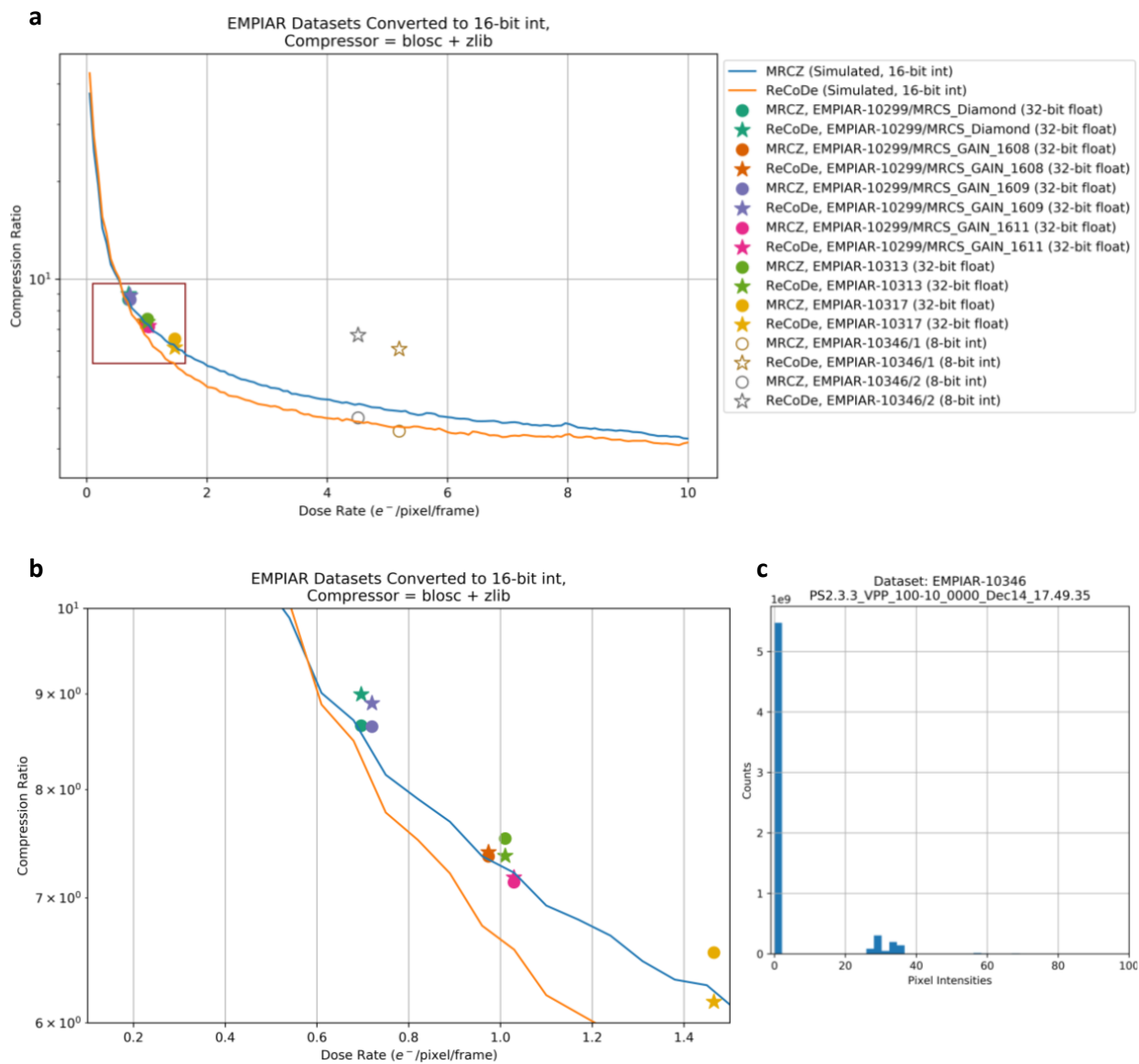

Supplementary Figure S7: Joint comparison of ReCoDe and MRCZ using simulated and EMPIAR datasets. (a) shows compression ratios achieved using ReCoDe and MRCZ on simulated and EMPIAR datasets. The simulated datasets had 16-bit unsigned integer data and for the purpose of standardisation, the EMPIAR datasets were also converted to 16-bit unsigned integer. The electron events per pixel follows a Poisson distribution in the simulated datasets. The underlying compression algorithms used for this comparison is Blosc + zlib (Deflate) for both MRCZ and ReCoDe. For the simulated data ReCoDe achieves higher compression ratios than MRCZ at lower dose rates (< 0.58 e/pix/frame). In case of EMPIAR datasets with 32-bit floating point data, ReCoDe gives slightly better compression than MRCZ up to dose rates <1 e/pix/frame. (b) shows the highlighted region of (a), containing the data point for the 32-bit floating point datasets, in greater detail. The crossover point for compression ratios achieved by ReCoDe and MRCZ can be seen to be approximately at a dose rate of 0.58 e/pix/frame for the simulated data. Whereas the crossover point for the EMPIAR datasets is closer to the dose rate of ~1.0 e/pix/frame, suggesting that electron events per pixel in the EMPIAR datasets deviate from the Poisson distribution and that these datasets are sparser than the simulated data. Note that the floating point data was converted to unsigned integer by normalising the pixel values to the 0-4096 range and rounding to the nearest integer. For the two datasets from EMPIAR-10346, that had 8-bit unsigned integer data, ReCoDe achieves much higher levels of compression than MRCZ, even though these datasets have a higher average dose rate. This is because these datasets contain sparse, high contrast frames.

(c) shows the distribution of pixel intensities for one of the EMPIAR-10346 dataset, acquired using a Volta phase plate. The histogram shows that the data is highly sparse with very high frequency of zeros and a small cluster of high count pixels, leading to an overall high average dose rate. Such sparse data enables ReCoDe to achieve higher compression ratios. Overall, these comparisons suggest that for various types of real datasets, ReCoDe can be an effective choice for data archival.

# Supplementary Figure S8: Shape and Size Distribution of Secondary Electron Puddles

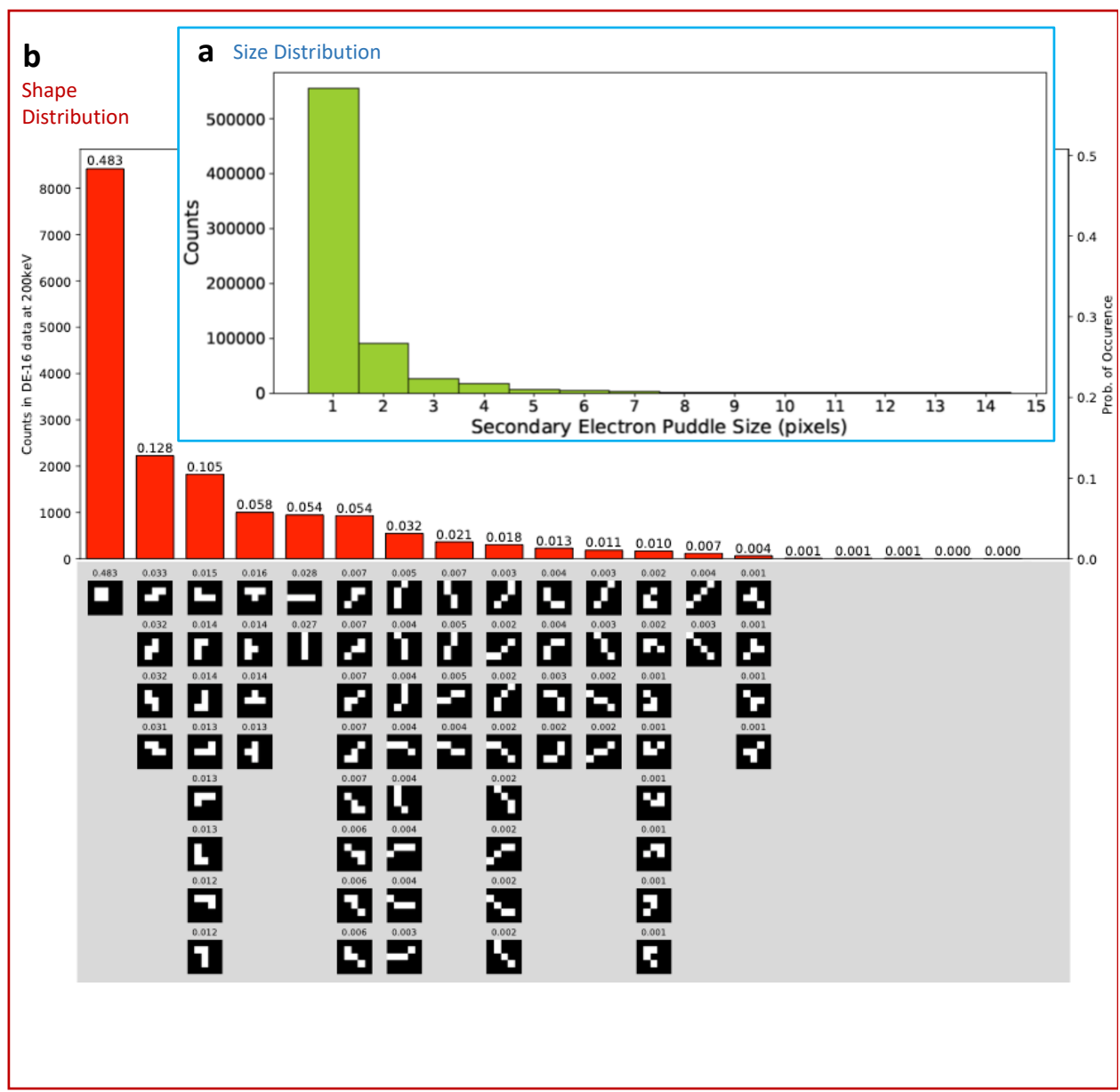

Supplementary Figure S8: Shape and size distribution of DE-16 secondary electron puddles. **(a)** The size distribution of 706,217 secondary electron puddles extracted from a DE-16 dataset acquired at a dose rate of 0.001 e-/pixel/sec. **(b)** Frequency of all possible rotational variants of electron puddles with an area of four pixels. The histogram tabulates the total number of counts for each unique four pixel puddle shape and its rotational variants and the corresponding probability of occurrence among the 17,419 four pixel sized electron puddles. Each bar in the plot represents one of the 16 unique shapes of four pixel sized puddles. Binarized representations of the puddle shapes, for orientations having occurrence probabilities higher than 0.001, are shown beneath. The shapes are sorted from left to right and then from top to bottom by their occurrence probabilities.

# Supplementary Figure S9: Compressibility of Representations of Centroids

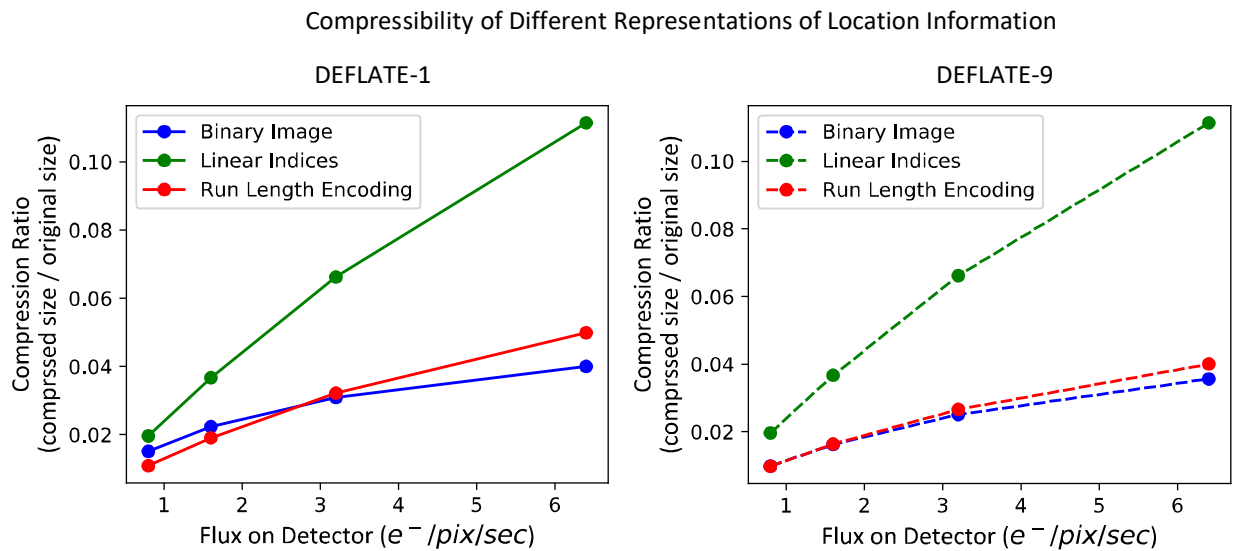

Supplementary Figure S9: Compressibility of three representations of centroids. Compression ratios due to three encodings of secondary electron puddle's locations when using **(a)** Deflate optimized for speed and **(b)** Deflate optimized for compression. In the Linear Indices encoding, a centroid's location is represented as a single 2n-bit linear index. In Run Length Encoding the linear indices are sorted and run-length encoded (RLE), since the ordering of centroids is inconsequential. In the Binary Image encoding, the centroids are represented as a binary image (similar to L4). RLE and binary image representations achieve comparable levels of compression and are much more compressible than linear indices. However, the binary image representation is more efficient than RLE, which requires a sorting. We therefore chose the binary image representation to encode the spatial information across all reduction levels of ReCoDe.

# Signal-Noise Calibration

## Supplementary Figure S10: DE-16 detector ADU distribution

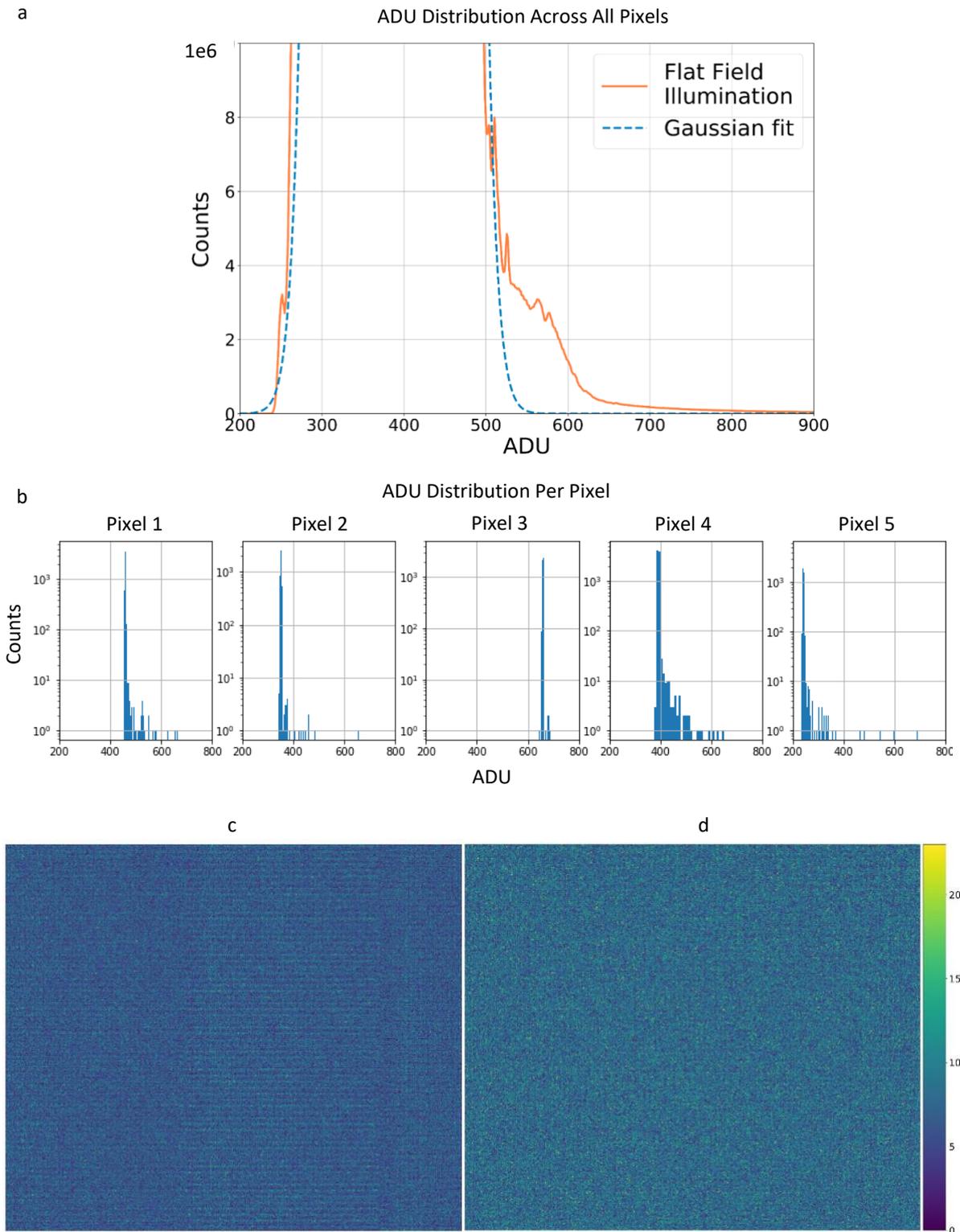

Supplementary Figure S10: DE-16 ADU distributions and adaptive calibration. **(a)** ADU distribution across all pixels in 10,000 frames of a low dose rate flat-field illumination frame-stack (0.001 e/pixel/frame), shows a smaller second distribution on the right due to actual electron events. **(b)** Visualizing the ADU distributions of individual pixels separately shows that pixels behave significantly differently from each other. **(c)** Using the same threshold for all pixels can severely bias counting. **(d)** The adaptive calibration used for DE-16 removes this bias.

## Supplementary Figure S11: On-the-fly Signal-Noise Calibration

*Definitions:*

$r$ = the tolerable false positive rate in detecting electron events

$R$ = stack of $n$ flat field illuminated frames

$X_{i,R} = (x_i^0, \ldots, x_i^n)$ be the values of the $i^{th}$ pixel across $n$ frames of $R$

*Signal-Noise Calibration Algorithm*:

1. Do dark subtraction: $F = R - \widetilde{D}$, where $\widetilde{D} = [d_i = median(X_{i,R})]$

2. Fit a normal distribution $\mathcal{N}(x_i \mid \mu_F, \sigma_F)$ to $F$

3. Compute the global threshold $\tau$, separating signal from noise, as $\tau = \mu_F + z * \sigma_F$, where z is given by:
$$r = P(Z \geq z) = \int_z^\infty \frac{1}{\sqrt{2\pi}} e^{\frac{-u^2}{2}} du$$

4. Select $m$ random patches of size $k \times k$, such that $k$ is odd and larger than the PSF's radius

5. Using $F$, identify outliers (signal), independently for each of the selected $mk^2$ pixels. Outliers in $X_{i,F}$: $\{x_i \mid x_i > \mu_{i,F} + z * \sigma_{i,F}\}$; assuming for the $i^{th}$ pixel $P(x_i) = \mathcal{N}(x_i \mid \mu_{i,F}, \sigma_{i,F})$

6. For each patch, compute $n_c$: the number of connected components formed by the outlier pixels originating at the central pixel of the patch

7. Compute the expected dose per pixel $\bar{n}_c$ as the mean $n_c$ across $m$ patches

8. For the $i^{th}$ pixel, compute gain $(g_i)$ as the median of the largest $\bar{n}_c$ values in $X_{i,F}$

9. The dark and gain corrected threshold for the $i^{th}$ pixel is:
$$\tau_i = \frac{\tau * \bar{g}}{g_i}$$

Supplementary Note S11: DE-16 On-the-fly signal-noise calibration algorithm for the DE-16 detector.

## Supplementary Note S12: Fine Signal-Noise Calibration for the DE-16 Detector

We performed a detailed calibration process to compare the average dose obtained against that from ReCoDe's fast calibration. The fine calibration procedure implements a common mode correction and an area thresholding step, in addition to the previous calibration steps. We implemented both calibration processes on flat-field illuminated datasets with different incident electron dosage. Comparing the counts obtained, we aim to provide an error estimate on the false positives and quantify the speed-accuracy trade-off.

Flat-field illuminated data frames were first collected at a sufficiently low dose rate that is comparable to the actual dose used for subsequent imaging. Next, we performed background noise subtraction by subtracting the median ADU value for each pixel. The median value is used since it provides a robust estimate of noise. With the dark subtracted dataset, we can proceed with a common-mode correction which reduces readout noise. To identify the common modes, the correlation between detector pixels must first be determined.

From a series of dark frames, collected in the absence of external illumination, we calculated the temporal correlations between pairs of pixels in a representative region of the detector. Next, we performed principal component analysis and transformed the pairwise correlation values, followed by K-means clustering. For the DE-16 detector, we identified correlations between pixels in every alternate column within each 1 by 256 block. The common-mode correction was therefore implemented block-wise per frame, with shared median values among the correlated pixels subtracted.

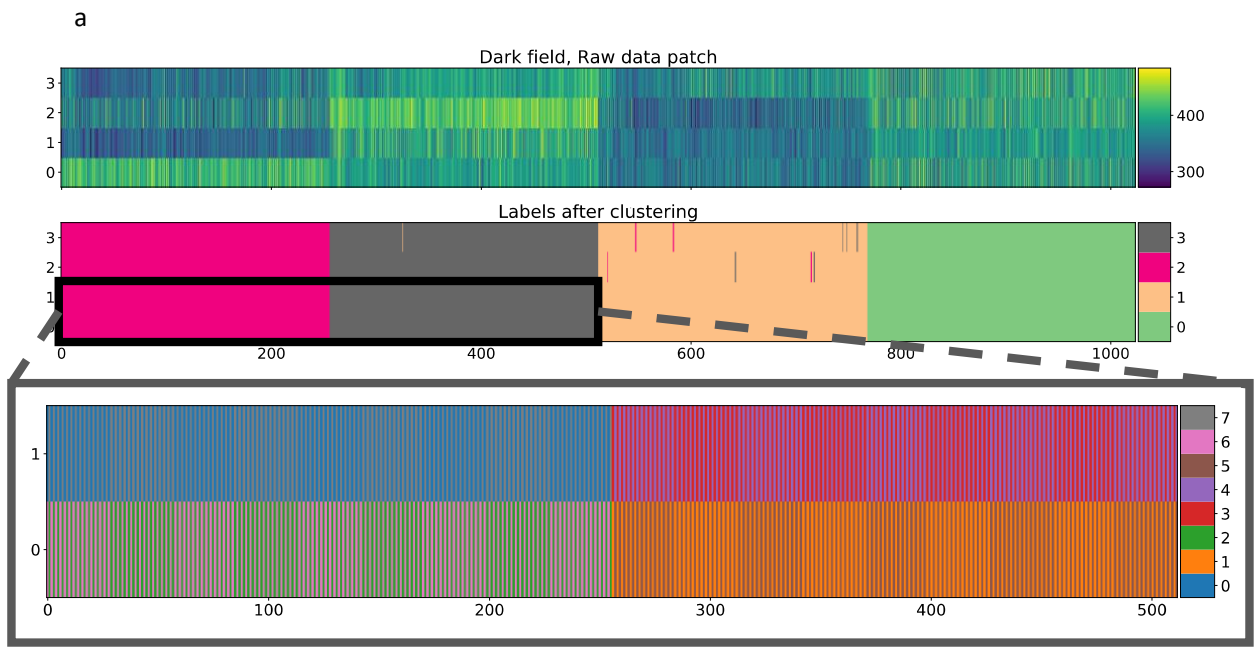

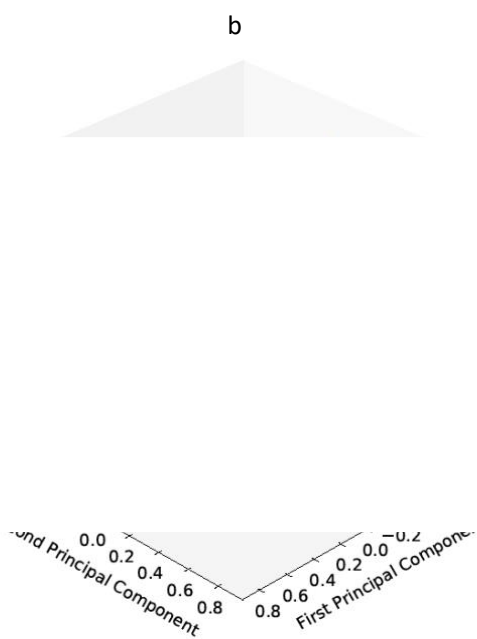

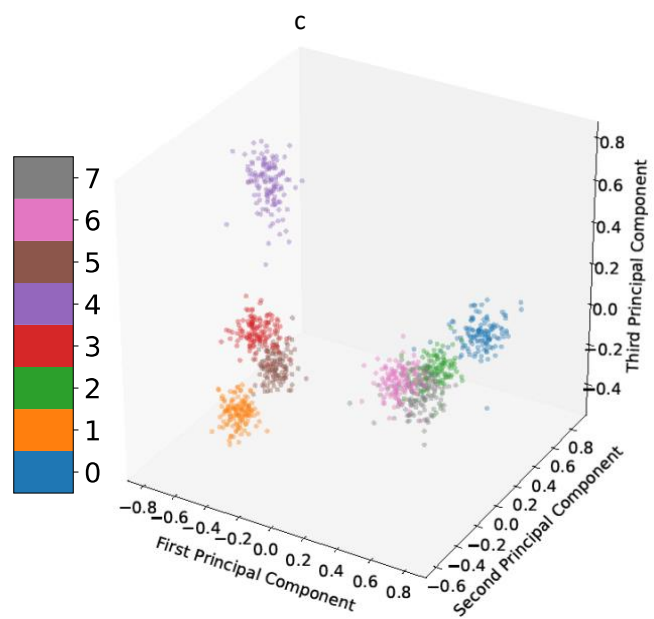

Supplementary Figure S12: Common mode correction for the DE-16 detector. (a) The topmost plot shows the response of 4 x 1024 pixel regions on the DE-16 detector in the absence of external illumination. From a series of dark frames, the pairwise temporal correlation between individual pixels, within the region was determined. The plot below shows the labels for the pixels after implementing principal component analysis (PCA), followed by k-means clustering on the pairwise correlation values. The pixels are grouped into four clusters, labelled with different colours. Pixels within each 4 x 256 block were observed to be grouped into the same cluster. The bottom plot shows the labels obtained for a 2 x 512 region spanning across two different blocks, after repeating the same process of PCA and k-means clustering with eight clusters. For each row of pixels, every alternate pixel across 256 columns was grouped into a single cluster. (b) k-means clustering of correlated pixels. The two 3D scatter plots show the transformed pairwise correlation values, for pixels in the 2 x 512 region, projected along the first, second and third principal components. Each data point represents a pixel, labelled according to its assigned cluster from k-means clustering. The same colour scheme is used to represent clusters as labels shown in the previous bottom plot. The plot on the left is a rotated view of the plot on the right, and together they show that all eight clusters are well separated from one another, with clear boundaries.

After the dark subtraction and common mode correction, we binned the ADU values of every pixel, across all frames, to construct a combined histogram. Assuming the dark non-signal values are distributed normally, we can determine an appropriate threshold to separate signal from noise. The threshold is chosen based on the maximum acceptable false positive rate, at the expense of losing true electron counts.

For a selected subset of data frames, binary maps were created where only pixels with ADU values higher than the noise threshold are marked as signal pixels. Connected signal pixels with neighbors of 2-connectivity were identified. Here, a minimal area constraint is implemented to reduce the false positive rate. Connected signal pixels must have an area larger than the predetermined threshold to be considered as an electron puddle. We finally count the total number of such puddles per frame and calculate the average electron dose per pixel.

With the estimated average electron dose per pixel per frame, we can calculate the expected total dose for each pixel in the selected subset of data frames. The estimated number of electron events each pixel registers across the subset is used to estimate the detector's gain response per pixel. With the assumption that events with larger ADU values are more likely to be signal events than noise, we can determine an ADU threshold to separate them. For each pixel, we first ranked all detected events by their ADU values from the largest to smallest. Since we expect the number of signal events to be at least as many as the estimated number of electron events determined from the average dose, we identify the event with its rank matching the expected number of signal events and the next event with a lower rank and ADU value. The ADU threshold is then determined from the mean ADU value of both events. All other events with ADU lower than the threshold will be labeled as noise.

We can now perform the final step in the fine calibration process. With the gain threshold determined for every pixel, the counting process is repeated to calculate the actual number of electron events after gain correction. Similar to before, binary maps were created from pixels with ADU values higher than their respective gain thresholds. These signal pixels are marked, and connected signal pixels with neighbors of 2-connectivity are identified as an electron puddle. The total number of electron puddles counted for each frame is the number of electron events for that frame. For the same selected subset of data frames, we now have a dark subtracted, common-mode and gain corrected average electron dose.

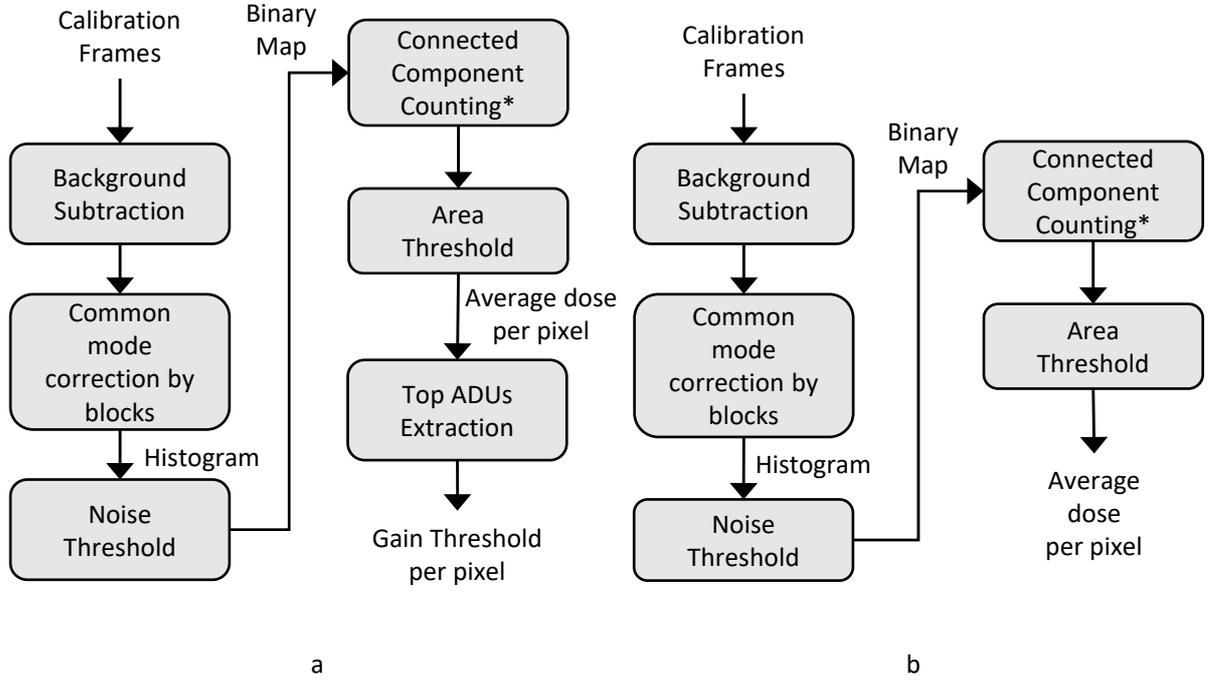

a  b

Supplementary Figure S13: Fine signal-noise calibration for DE-16 data. A variant of the on-the-fly calibration procedure described in Supplementary Figure S9 is presented here. We refer to this as a "fine calibration", as it incorporates common mode correction as well as an additional area threshold. After background noise subtraction across the calibration frames, common mode correction was performed through identifying correlated pixels and subtracting their shared median values. A combined histogram of all corrected ADU values was constructed, from which an appropriate noise threshold was determined to produce a binary map which differentiated signal from noise. From a selected subset of frames, connected signal pixels were identified and counted to estimate the number of clusters per frame. Only clusters of connected pixels which satisfy a minimal area threshold were considered to be electron puddles. The total number of such puddles provides an estimate on the average electron dose per frame. For each pixel, all detected events are then ranked by their ADU values from the largest to smallest. With the expected total number of electron counts across the subset of frames, the event at this rank and the next event with a lower rank are selected. The gain threshold for each pixel is then determined from the mean ADU value of both events.

The experiment conducted to compare ReCoDe's on-the-fly calibration and the fine calibration was as follows. A series of flat-field illuminated data frames were collected with a specified electron beam magnification. The dataset was processed using both calibration methods and the average electron dose was recorded. The same process was then repeated at varying electron beam magnifications since the incident electron dose scales inversely with the square of beam magnification. The change in electron counts against the incident dose, which we fitted to a logarithmic model Eqn. (1), with FP as the number of false positives, m as the power term for beam magnification and $\rho_0$ as the counts at the asymptotic limit of zero magnification, allows us to compare between the two calibration methods.

$$ln(\boldsymbol{Counts}) = ln(FP) + ln(1 + \frac{\rho_0 A}{FP + Mag^m}) \qquad (1)$$

We also collected datasets using DE-16's electron counting mode, following the same electron beam magnifications. These counts were also fitted to the same model to compare the differences between the three methods. The curves fitted and the corresponding values for the fitting parameters are stated in Supplementary Table S3.

The fine calibration routine resulted in an approximate 30% decrease in identified electron counts, compared to ReCoDe's fast calibration. As shown in Supplementary Figure S14, the main contributing factor is the area constraint imposed when identifying valid events during the estimation of the average electron dose, which was subsequently used to calibrate each pixel's gain response. The correction for common-mode in the dataset did not significantly affect the final estimated electron counts. The difference in counts was likely due to events that were incorrectly identified as a signal during the initial counting when estimating the average dose per frame. These results suggest that the on-the-fly calibration routine might be more lenient, keeping a larger number of electron events at the cost of misidentifying some noise fluctuations as electron events. Nevertheless, a more stringent recalibration process can always be implemented post reduction for L1 to L4 reduction, if required.

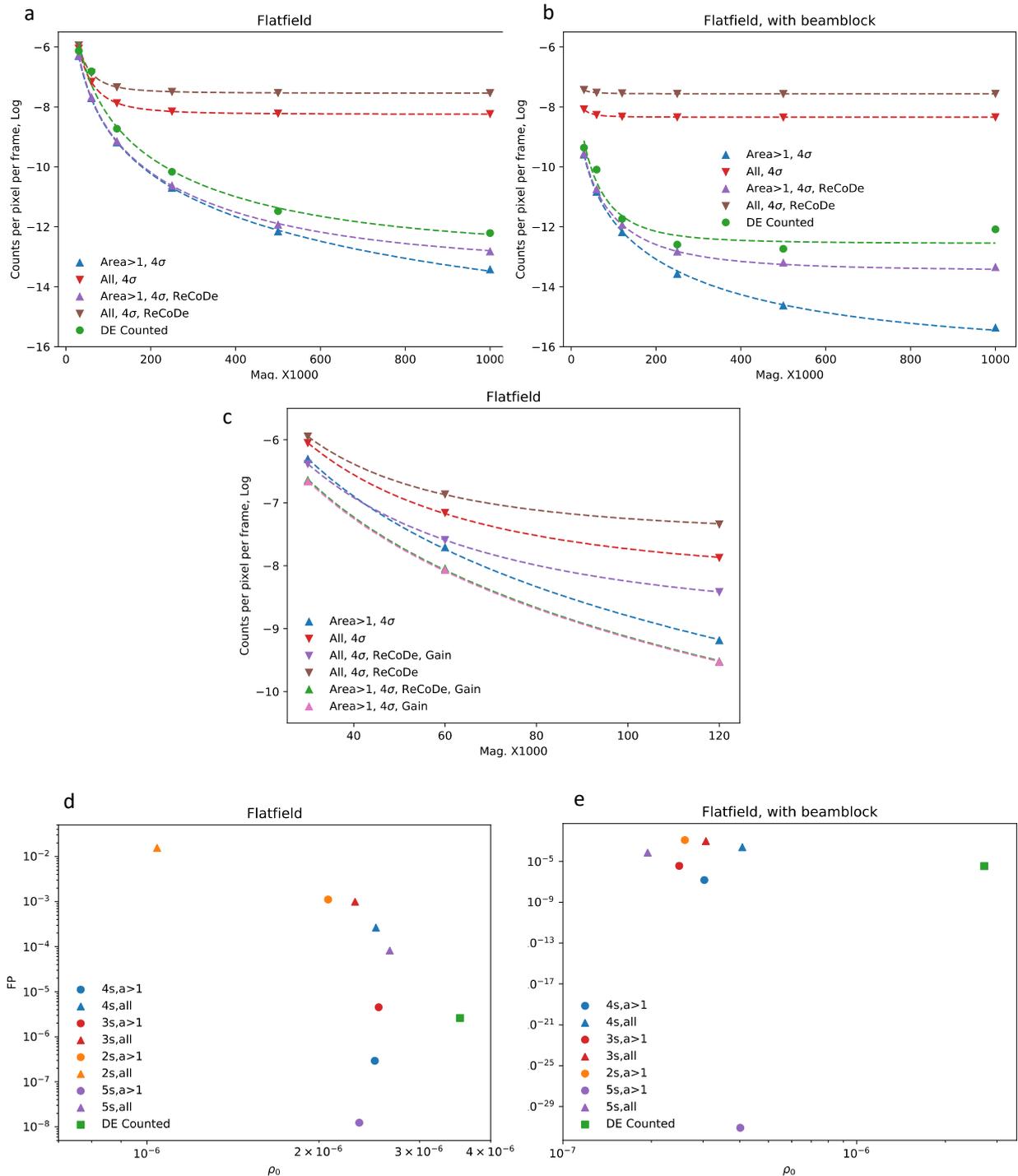

Supplementary Figure S14: Evaluation of on-the-fly and fine calibrations. (a) and (b) Estimated electron counts for varying electron beam magnifications. The average electron dose, with an intensity threshold of 4 standard

deviations from the mean noise ADU value, was estimated for different magnifications of electron beam, with and without an area constraint of more than 1 pixel on the electron puddle size. The average electron count, including those obtained through DE-16's electron counting, follows an inverse relationship with the square of the beam magnification. The counts estimated with the area constraint matches the DE counts much closely, compared to without the constraint. Comparatively, estimated average electron dose from ReCoDe's fast calibration follows the same trend only when the magnification was low, diverging when the magnification increased. With the beamblock, DE counts approached an asymptote at high magnification while estimated counts with the area constraint continued to follow a decreasing trend similar to before. Estimated average electron dose with ReCoDe's fast calibration similarly plateaus at high magnification, albeit having a lower asymptotic limit than DE counts. The fitted curves and the corresponding fitting parameters are listed in Supplementary Table S3. (c) Electron counts after per pixel gain calibration, for varying electron beam magnification. After calibrating for each pixel's gain response, following the fine calibration routine resulted in a ~30% decrease in identified electron events, as compared to ReCoDe's fast calibration routine. The main factor contributing to the reduction in counts is the area constraint imposed on each event, which can significantly reduce the number of false positives. Comparatively, correcting for the common mode did not result in a significant change in the number of counts. These results suggest that ReCoDe's calibration routine is more lenient, allowing a number of noise fluctuations to be misidentified as electron events. Nevertheless, a more stringent recalibration process can be implemented post reduction for L1 to L3 reduced data, if required. This is definitely preferable compared to the alternative where strict constraints are implemented during the early calibration process of the reduction and compression pipeline, which will make it extremely difficult to recover potential events that have been removed. This suggests that there could be further room for data compression if a lower true positive rate can be accepted. (d) and (e) Number of false positives and electron counts at the asymptotic limit of zero magnification, $\rho_0$. Imposing an area constraint significantly reduced the number of false positives in the estimation of average electron dose, as compared to accepting all identified events including those which only occupy a single pixel. As the intensity threshold increases, the number of false positives decreases consequently, since the number of misidentified events contributed by noise decreases. Unfortunately, this comes at the cost of sacrificing true electron events with low energies, as reflected by the parameter $\rho_0$, the theoretical maximum for the number of true electron events. Therefore, an appropriate threshold is required to achieve the best compromise, allowing for high true positive rates while maintaining low false positive rates.

Supplementary Table S3: Fitting params used for evaluating on-the-fly and fine calibrations

$$\ln(\boldsymbol{Counts}) = \ln(\boldsymbol{FP}) + \ln\left(\mathbf{1} + \frac{\rho_0 A}{FP \times Mag^m}\right)$$

A: 921600 px

|  | FP | m | $\rho_0$ |
|---|---|---|---|
| Area > 1, $4\sigma$ | $1.3(4) \times 10^{-7}$ | $2.080(2)$ | $2.32(2) \times 10^{-6}$ |
| All area, $4\sigma$ | $2.617(5) \times 10^{-4}$ | $2.060(4)$ | $2.49(3) \times 10^{-6}$ |
| Area > 1, $4\sigma$, ReCoDe | $1.52(8) \times 10^{-6}$ | $2.09(1)$ | $2.5(2) \times 10^{-6}$ |
| All area, $4\sigma$, ReCoDe | $5.30(1) \times 10^{-4}$ | $2.04(1)$ | $2.3(1) \times 10^{-6}$ |
| DE Counting | $3(2) \times 10^{-6}$ | $2.1(2)$ | $4(3) \times 10^{-6}$ |

With beamblock
A: 178161 px

|  | FP | m | $\rho_0$ |
|---|---|---|---|
| Area > 1, $4\sigma$ | $1.0(3) \times 10^{-7}$ | $1.87(2)$ | $2.2(2) \times 10^{-7}$ |
| All area, $4\sigma$ | $2.378(4) \times 10^{-4}$ | $1.90(9)$ | $2.6(8) \times 10^{-7}$ |
| Area > 1, $4\sigma$, ReCoDe | $1.39(4) \times 10^{-6}$ | $1.87(3)$ | $2.3(3) \times 10^{-7}$ |
| All area, $4\sigma$, ReCoDe | $5.20(3) \times 10^{-4}$ | $1.8(4)$ | $2(2) \times 10^{-7}$ |
| DE Counting | $4(1) \times 10^{-6}$ | $2.2(6)$ | $3(7) \times 10^{-6}$ |

# Supplementary Note S13: Estimating Backscattering

We estimated backscattering ratio by comparing simulated primary and backscattered electron events with actual data. The simulation model employed contains three main parameters, namely the total number of events per frame, the ratio of primary-to-backscattered events and the distribution model for nearest distance between a backscattered and a primary event. For each simulation, the total number of events per frame is fixed following the event counts obtained experimentally. The ratio of primary-to-backscattered events then determines the number of simulated backscattered events in each frame. The distance between a backscattered event and its nearest primary event is assumed to follow an exponential distribution, with the sum of the location parameter and the inverse of the lambda parameter as the mean nearest-neighbour distance for each backscattered event. Assuming primary events are uniformly scattered across the entire frame, a subset of them are randomly selected as the neighbours of backscattered events. Each backscattered event is then randomly placed beside its primary counterpart, at a distance sampled from the exponential distribution. The nearest-neighbour distances are then calculated for all simulated events, and tabulated into a histogram to be compared with the nearest-neighbour distance histogram from the actual electron events. A constraint of nearest-neighbour distances to be larger than two pixels was applied on the simulated events to match the experimental events, propagated from the minimum area constraint on electron puddles. A two-sample Kolmogorov–Smirnov (K-S) test between distributions of nearest neighbour distance for simulated and experimental data to obtain the D-statistic which quantifies the difference between two histograms. The minimum value for D-statistic among 100 repeated simulations for each parameter pair are plotted, and the optimal values are in the range of 8 to 9 (~8.6) for the primary-to-backscattered ratio, and 6 to 7 pixels (~6.4) in average nearest-neighbour distance for a backscattered event.

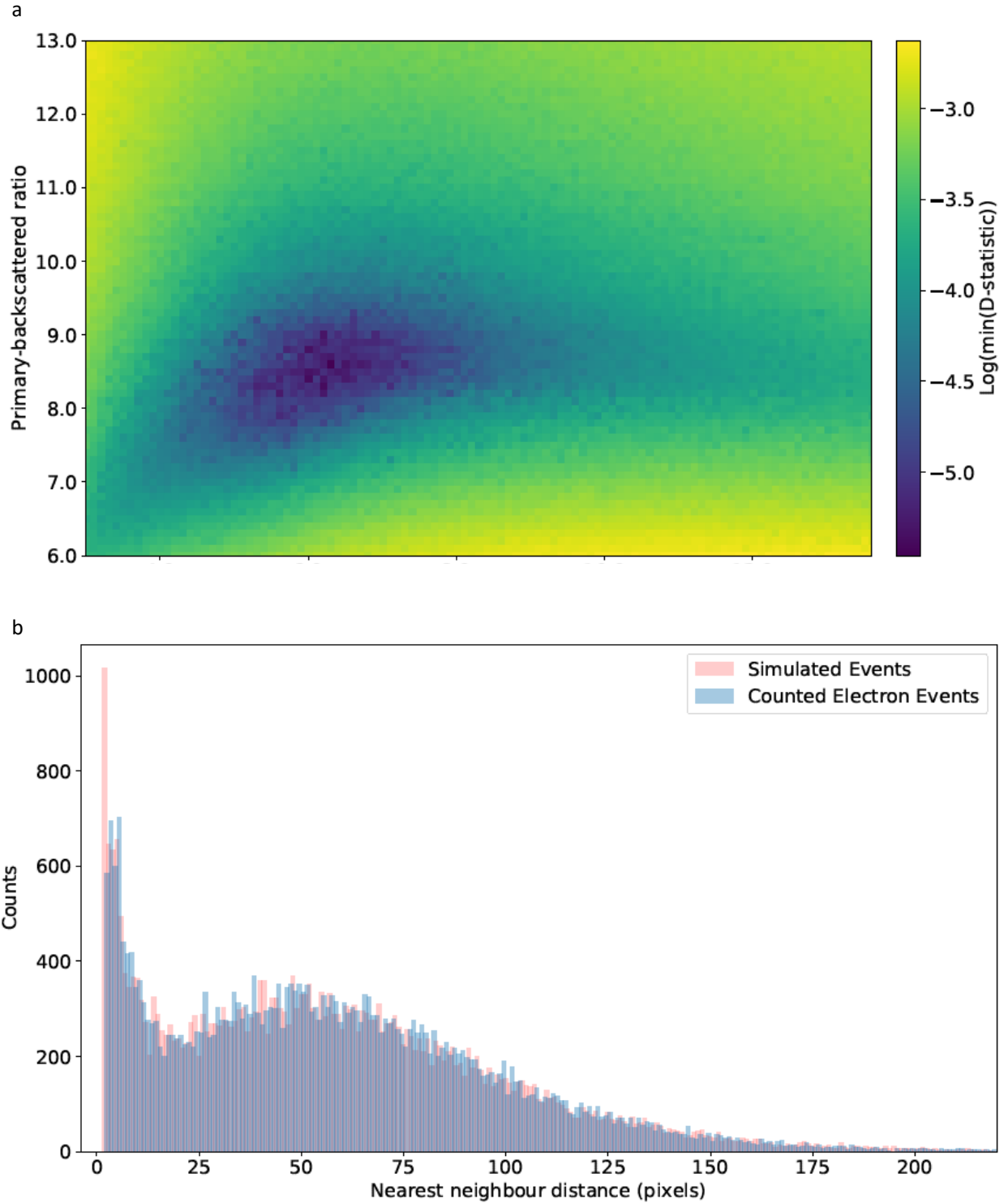

Supplementary Figure S15: Estimating backscattering ratio by modelling primary and backscattered electron events with simulations. (a) The minimum value for D-statistic among 100 repeated simulations for each parameter pair. The optimal values are in the range of 8 to 9 (~8.6) for the primary-to-backscattered ratio, and 6 to 7 pixels (~6.4) in average nearest-neighbour distance for a backscattered event. (b) Histograms of nearest neighbour distances between electron events for both simulated and experimental counted primary and backscattered events. The simulated primary and secondary events are generated following the parameters, a ratio of 8.7 for primary to backscattered events per frame, and 6.4 pixels for the mean nearest-neighbour distance for a backscattered event. For both simulated and experimental dataset, the nearest-neighbour distances are first calculated for all events in each frame. The total number of counts for each distance are then accumulated across all frames, and finally

tabulated into a histogram for comparison. These parameter values gave a combined q-value, following Fisher's method with 100 simulations, of more than a significance level of 0.01, indicating that both histograms were sampled from the same distribution. Comparing the simulated and experimental histograms, K-S test gave a D-statistic of 0.00607, with a corresponding p-value of 0.587, after the constraint of minimum nearest neighbour distance of two pixels was applied.

# References

1. Amdahl, G. M. Validity of the single processor approach to achieving large scale computing capabilities. *Proceedings of the April 18-20, 1967, spring joint* (1967).